\documentclass[12pt,preprint]{aastex}

\usepackage{amsmath,amssymb}

\newcommand{\sub}[1]{_{\text{#1}}}
\newcommand{\super}[1]{^{\text{#1}}}

\slugcomment{\apj \ in press}

\shorttitle{Dust Growth and Settling in Protoplanetary Disks I.}
\shortauthors{Tanaka et al.}

\begin{document}

\title{Dust Growth and Settling in Protoplanetary Disks\\
and Disk Spectral Energy Distributions:\\
I. Laminar Disks}

\author{Hidekazu Tanaka, Youhei Himeno, and Shigeru Ida}
\affil{Department of Earth and Planetary Sciences, 
Tokyo institute of Technology\\
 Tokyo 152-8551, Japan}
\email{hidekazu@geo.titech.ac.jp}

\begin{abstract}
Dust growth and settling considerably affect the spectral energy 
distributions (SEDs) of protoplanetary disks.
We investigated dust growth and settling in 
protoplanetary disks through numerical simulations to examine
time-evolution of the disk optical thickness and SEDs.
In the present paper, we considered laminar disks as a first step
of a series of papers.
As a result of dust growth and settling,
a dust layer forms around the mid-plane of a gaseous disk.
After the formation of the dust layer,
small dust grains remain floating above the layer.
Although the surface density of the floating small grains
is much less than that of the dust layer,
they govern the disk optical thickness and the emission.
Size distributions of the floating grains obtained from 
numerical simulations are well-described by an universal power-law 
distribution, which is independent of the disk temperature, the disk 
surface density, the radial position in the disk, etc. 
The floating small grains settle onto the dust layer in a long time scale
compared with the formation of the dust layer.
Typically, it takes $10^6$yr for micron-sized grains. 
Rapid grain growth in the inner part of disks makes
the radial distribution of the disk optical thickness
less steep than that of the disk surface density, $\varSigma$.
For disks with $\varSigma \propto R^{-3/2}$, 
the radial distribution of the optical thickness is almost flat for all 
wavelengths at $t\lesssim 10^6$yr.
At $t > 10^6$yr, the optical thickness of the inner disk ($\lesssim $ a few
AU) almost vanishes, which may correspond to
disk inner holes observed by Spitzer Space Telescope.
Furthermore, we examined time-evolution of disk SEDs,
using our numerical results and the two-layer model.
The grain growth and settling decrease the magnitude of the SEDs
especially at $\lambda \ge 100\mu$m. Our results indicate that
grain growth and settling can explain 
the decrease in observed energy fluxes at millimeter/sub-millimeter 
wavelengths with time scales of $10^{6-7}$yr without depletion of the disks.
\end{abstract}

\section{INTRODUCTION}\label{sec:intro}

Circumstellar disks surrounding T Tauri and Herbig Ae/Be stars are thought to
the birth sites of planetary systems and they are called protoplanetary disks.
The physical properties of the protoplanetary disks have been inferred from 
analyses of their spectral energy distributions (SEDs).
Analyses of observations at millimeter/sub-millimeter wavelengths
suggest that the disks have masses of 0.001-0.1 $M_{\odot}$ 
and their life time is $\sim 10^{6-7}$yr
(e.g., Strom et al.\ 1989; Beckwith et al.\ 1990; Wyatt et al. 2003).
 
Recent observations with high-resolution at various wavelengths
gave us much information on disk spatial structure 
(e.g., Grady et al. 1999; Fukagawa et al. 2004;
Mundy et al. 1996; Kitamura et al. 2002; Eisner et al. 2003).
For example, Kitamura et al.\ (2002) 
derived disk outer radii and surface density distributions
from results of their image survey of 13 T Tauri disks and SED data.
Their results show that the radial power-law index of the surface density 
$\varSigma(R)$ ranges from 0 to -1, which is less steep than -1.5 in the
minimum mass solar nebula (MMSN) (Hayashi 1981).
Because the obtained surface densities at 100 AU is almost consistent with 
the MMSN, the small radial power-law index of $\varSigma$ means too small 
surface density at $R \le 10$AU for planet formation in those disks. 

In the most analyses of disk observations including Kitamura et al.\ (2002),
dust growth have not been taken into account.
However, the grain growth is indicated by the variety of silicate 
feature at 10$\mu$m observed in T Tauri disks 
(Przygodda et al.~2003; Honda et al.~2003; Meeus et al.~2003).
Dust growth usually decreases the opacity and the optical thickness of disks.
If grains grow considerably in the observed disks, the estimated disk 
mass mentioned above would be underestimated 
(D'Alessio et al. 2001; Natta et al. 2004).
Furthermore, dust growth is generally fast in the inner disk region compared 
with the outer region (Weidenschilling 1980; Nakagawa et al. 1981). 
Rapid decrease in the optical thickness in the inner region would make the 
radial distribution of the optical thickness less steep than that of the 
surface density. The rapid dust growth in inner regions may also be responsible
to inner holes identified by Calvet et al. (2002) in TW Hydra and by
Bouwman et al. (2003) in HD1000546.
Furthermore, the Spitzer Space Telescope started to provide lots of new data 
on disk inner holes (Forrest at al. 2004; Meyer et al. 2004; 
Uchida et al. 2004). 
Hence, dust growth should be taken into account in the 
analyses of disk observations.

Recent models of protoplanetary disks succeeded in accounting for most 
properties in the observed SEDs (e.g., Chiang \& Goldreich 1997; 
Chiang et al. 2001; D'Alessio et al. 2001; Dullemond et al. 2001; 
Dullemond \& Dominik 2004a). Although they also examined the effect of dust 
growth, they a priori adopted a single power-law size distribution for a whole 
disk and assumed that the grain number at each size is proportional to the 
gas density. 
However, the size distribution would be considerably different at 
each position because of a large variety of the growth time in a whole disk,
as mentioned above.  Furthermore, dust settling would change the vertical 
distribution of dust grain in disks. 
Dullemond \& Dominik (2004b) showed the importance of
dust settling at the self-shadowing process in passive disks.
To derive information correctly from observations,
the more realistic model of dust growth and settling should be used.

Dust growth and settling in protoplanetary disks have been investigated by 
many 
authors for both laminar disks (e.g., Safronov 1969; Weidenschilling 1980, 
1997; 
Nakagawa et al.\ 1981, 1986) and turbulent disks (e.g., Weidenschilling 1984;
Mizuno et al.\ 1988; Schmitt et al. 1997). In laminar disks or those with 
sufficiently weak turbulence, dust grains grow to mm-size when they settle 
to the mid-plane of the gaseous disk. As a result of dust settling, 
a dust-rich layer is formed near the mid-plane. 
In the dust layer, a large number of planetesimals are thought
to form as the first step of planetary formation (e.g., Goldreich \& Ward 1973;
Weidenschilling 1997; Youdin \& Shu 2002). Because most previous studies 
focused on the dust layer and the planetesimal formation, they did not examine
in detail a relatively small amount of micron-sized grains that remain 
floating after the dust layer formation. However, as suggested by 
Nakagawa et al (1981) and some authors, the floating small grains would 
play an important role in the radiative transfer in protoplanetary disks 
even though their total mass is small.  
The change in SEDs due to settling of such floating grains was examined by 
Miyake \& Nakagawa (1995) and by Dullemond \& Dominik (2004b).
However, they did not consider settling coupled with growth.
The variation in the disk optical thickness due to dust growth
was also evaluated in some studies on dust growth and settling
(Weidenschilling 1984, 1997; Weidenschilling \& Cuzzi 1993; 
Schmitt et al. 1997).
However, these evaluations were done only in short-term calculations 
($\lesssim 100$yr) or only at limited radial positions (i.e., 1 and 30 AU).
In order to take into account dust growth in analyses of disk observations,
it is necessary to clarify the general property in the long-term evolution of 
floating grains and disk optical depth in various disks.

In this series of papers, we investigate dust growth and settling
in protoplanetary disks, focusing on floating small grains
rather than large grains in dust layers. 
We perform numerical simulations of dust growth and settling
for various disk models
to obtain the general property of the spatial and size distribution of dust grains.
Using the obtained spatial and size distribution of grains,
we further examine time-evolution of the disk optical thickness and SEDs
due to dust growth and settling.  Our results show that, because of relatively rapid dust growth in the inner disk 
region, the disk optical thickness decreases more rapidly in the inner region
than in the outer. This makes the radial distribution of the optical thickness 
less steep, as expected above. 
The dust emission at millimeter/sub-millimeter wavelengths declines 
on the time scale $\sim 10^6$ yr even without any disk depletion.
Therefore, the observed decay of dust emission does not 
necessarily indicate disk depletion. 

In the present paper, we consider laminar disks as the first step
of this series of papers. As shown by Dullemond \& Dominik (2004b),
vertical stirring by turbulence is effective on dust settling
even in weakly turbulent disks with the parameter $\alpha =10^{-4}$
for the disk viscosity (Shakura \& Sunyaev 1973).
Furthermore, since turbulence would play a crucial role in the issue of 
planetesimal formation in a dust layer by self-gravitational 
instability, it cannot be neglected on that issue.
However, on the disk optical thickness,
essential ingredients would be included in laminar disk cases. 
Thus the calculations with laminar disks would give deep insights into 
time-evolution of the optical thickness and disk SEDs, 
although the time scale can be affected by turbulence.
Hence, the case in laminar disks should be firstly examined 
as a simple case. The case in turbulent disks will be investigated in the 
subsequent paper. 

In the next section, we describe basic equations and our numerical model 
of growth and settling. 
We show numerical results for various disk models in Section 3.
We also derive an analytical solution of the grain size distribution, 
which agrees well with numerical results.
In Section 4, we examine time-evolution of the disk optical thickness
and SEDs due to dust growth and settling. In the calculation of SEDs, we used 
the two-layer model proposed by Chiang \& Goldreich (1997) and refined by
Chiang et al. (2001).
In the last section, we summarize our results and discuss the effect of 
turbulence on dust growth and settling.
\section{EQUATIONS DESCRIBING DUST GROWTH AND SETTLING}\label{sec2}

\subsection{Disk Model}\label{subsec:diskmodel}

We consider a passive protoplanetary disk with no turbulence around a central
stars with one solar mass ($M_*=1M_\odot$). 
It is assumed that the disk has an inner edge at 0.1 AU and an outer edge 
at 150 AU.
To describe the disk, we use a cylindrical coordinate system $(R, \phi, z)$
of which the origin is located at the central star. The $z$-axis
coincide with the rotation axis of the disk.
In numerical simulations of dust growth and settling,
we assume that the gaseous disk is isothermal in the vertical direction
(i.e., in the $z$-direction) for simplicity.
As seen in the two-layer model of passive disks, however, the surface layer 
directly irradiated by the central star is much hotter than the disk interior.
Nevertheless, our modeling of a vertical isothermal disk would be valid
because dust growth and settling in laminar disks are little 
affected by the disk temperature, as seen in the next section.
In the evaluation of SEDs in Section 4, of course, the vertical temperature
distribution will be taken into account with the two-layer model.

Under this assumption, the vertical density distribution $\rho\sub{g}(z)$
of the gas component is given in the hydrostatic equilibrium by
\begin{equation}
\rho\sub{g}(z) = \frac{\varSigma\sub{g}}{\sqrt{2\pi} h_p} 
                 \exp\left(-\frac{z^2}{2h_p^2}\right),
\label{eq:dens}
\end{equation}
where $\varSigma\sub{g}$ is the gas surface density and $h_p$ is 
the pressure scale height of the disk given by $c\sub{s}/\varOmega\sub{K}$.
Furthermore, $\varOmega\sub{K}$ $=\sqrt{GM_\odot/R^3}$ is the Keplerian 
angular velocity and $c\sub{s}=\sqrt{ k\sub{B} T/(\mu m\sub{u})}$
is the isothermal sound speed, where $G$, $k\sub{B}$, and 
$m\sub{u}$ are the gravitational constant, Boltzmann's constant, and the 
atomic mass unit, respectively; and $T$ and $\mu$ are the disk temperature 
and the mean molecular weight, respectively. We set $\mu = 2.34$.

Dust growth and settling change the disk temperature.
Thus, exactly speaking, we need to solve coupled equations
of dust growth, settling and the disk temperature.
However, since a change in the disk temperature has a little effect
on dust growth and settling, we use a tentative disk temperature 
for simulations of dust growth and settling.
The tentative disk temperature is calculated in the same manner as the 
calculation of the temperature of the disk interior of the two-layer model 
(i.e., using equations [\ref{eq:Ti}] and [\ref{eq:angle}] in the appendix) 
but, in this calculation, we assumed that the disk is optically thick 
($\tau\sub{P,i}\gg 1$) and that the height of the surface layer, 
$z\sub{s}$, is equal to $\sqrt{2} h_p$.
Note that we will correctly evaluate the temperatures of the disk interior
and the surface layer according to the two-layer model
in the calculation of the disk SED (Section 4).
The deviation of the tentative temperature
from that of the disk interior reevaluated in Section 4 is within 30\%.

As for the surface density distribution of gaseous disks,
we adopt a power-law distribution of
\begin{equation}
\varSigma\sub{g}(R) = \varSigma\sub{g1}
                      \left( \frac{R}{1 \mbox{AU}} \right)^{-s}.
\label{eq:sdens_g}
\end{equation}
The gas surface density at 1AU, $\varSigma\sub{g1}$, and the power-law index 
$s$ are parameters. In the standard case of the present paper, we adopt the 
values of the MMSN, that is, 
$\varSigma\sub{g1} = 1700$gcm$^{-3}$ and $s=1.5$.
The dust-to-gas ratio $\zeta\sub{d}$ and the solid material density 
$\rho\sub{s}$ are set to be 
0.014 and 1.4gcm$^{-3}$, respectively in the standard case.
These values are consistent 
with the solar abundance when H$_2$O ice is included in grains
(Pollack et al.1994).
We also examine cases with a different value of $\zeta\sub{d}$ or 
$\rho\sub{s}$.
At $T>150K$,  H$_2$O evaporates and we should decrease the dust-gas ratio
and increase the solid material density.
However, we neglected these changes in the simulations of dust growth and 
settling for simplicity.

\subsection{Dust Growth and Settling}\label{subsec:growth}

To describe dust growth and settling, we adopted the basic equations 
used by Nakagawa et al.\ (1981).
Let $n(m,z) dm$ denote the number density of grains with masses 
between $m$ and $m + dm$ at a height $z$. Thus $n(m,z)$ describes 
the distribution of grains with respect to $m$ (or the size) 
and $z$. The statistical coagulation equation is used to 
describe evolution of the mass distribution due to collisional growth.
To describe dust settling as well as growth, a vertical advection term 
is added to the coagulation equation. That is, the following equation is used:
\begin{eqnarray}
&&\frac{\partial}{\partial t} n(m,z)
+ \frac{\partial}{\partial z} \left[ V_z(m,z) n(m,z) \right]
\nonumber \\
&& \quad = \frac{1}{2} \int_0^m K_{m',m - m'} n(m',z) n(m - m',z) dm' 
\nonumber \\
&&\quad \quad - n(m,z) \int_0^{\infty} K_{m,m'} n(m',z) dm',
\label{eq:coagsedi1}
\end{eqnarray}
where the kernel, $K_{m',m - m'}$, is related to the coalescence rate 
between grains with masses $m$ and $m - m'$, and
$V_z$ is the settling velocity of dust grains.
In the above, the first term on the right-hand side indicates 
the formation of grains of the mass $m$ by collisions between the 
smaller grains and the second term represents the consumption of 
grains with the mass $m$ due to collisions with other grains.
The second term on the left-hand side is the advection one, which 
describes the vertical mass transport. In equation~(\ref{eq:coagsedi1}), 
we neglected the advection term due to the radial mass transport 
because it is effective only for relatively large grains that settle down to
the dust layer.
We also neglected the diffusion term due to turbulence
since laminar disks are considered.

The kernel $K_{m,m'}$ is given by $p\sub{s} \sigma_{m,m'} \Delta V_{m,m'}$, 
where $\sigma_{m,m'}$ is the collisional cross section between grains with
masses  $m$ and $m'$, $p\sub{s}$ is the sticking probability, and 
$\Delta V_{m,m'}$ is the relative velocity between grains with $m$ and $m'$. 
For dust grains, $\sigma_{m,m'}$ is equal to the geometrical cross section. 
We assume spherical grains in the present study. Then we obtain the cross 
section $\sigma_{m,m'}$ as $\pi \left[a(m) + a(m')\right]^2$, where $a(m)$ 
is the radius of grains with $m$. The relative velocity $\Delta V_{m,m'}$ 
will be explained in the next subsection.
The sticking probability $p\sub{s}$ is difficult to be determined
because it depends on its size, shape, material properties 
(e.g., Weidenschilling \& Cuzzi 1993). 
In the present study, we regard $p\sub{s}$ as a parameter and 
assume that it is independent of the grain size.

\subsection{Motion of dust grains in protoplanetary disks}
\label{subsec:motion}

In a laminar disk, grains settle toward its mid-plane. 
The settling velocity of each grain is given by the terminal velocity, which
is determined by the balance between the gas drag force and the vertical 
component of the central star's gravity.
As well as such a settling velocity, Nakagawa et al. (1986) also calculate 
the horizontal components of the grain velocity,
assuming an system of equal-sized grains.
Here, we describe motion of grains with an arbitrary size distribution

We first explain the gas drag forces on grains.
For a spherical grain with a mass $m$, 
the gas drag force is given by
\begin{equation}
{\boldsymbol{F}}\sub{D} = - m A(m) \rho\sub{g}
({\boldsymbol{U}}(m)-{\boldsymbol{u}})
\label{eq:gas_drag3}
\end{equation}
and the coefficient $A$ is given by
\begin{equation}
A(m) = \left\{
\begin{array}{lc}
\displaystyle
\sqrt{\frac{8}{\pi}} \frac{c\sub{s}}{\rho\sub{s} a(m)} & 
\quad (a \le \frac{3}{2} l), \\[4mm]
\displaystyle
\sqrt{\frac{8}{\pi}} \frac{3 l c\sub{s}}{2 \rho\sub{s} a(m)^2} &
\quad (a > \frac{3}{2} l),  
\end{array}
\right.
\label{eq:gas_drag4}
\end{equation}
where ${\boldsymbol{u}}$ and ${\boldsymbol{U}}(m)$ are the velocities of gas 
and the grain, respectively and $l$ is the mean free path.
In equation~(\ref{eq:gas_drag4}), the first case corresponds
to Epstein's law and the second is Stokes' law.
The mean free path is given by 
$l = \mu m\sub{u}/(\rho\sub{g} \sigma\sub{mol})$, where $\sigma\sub{mol}$ 
is the molecular collision cross section given by 
$2 \times 10^{-15}$cm$^2$ (Chapman \& Cowling 1970).
At 1AU and 10AU on the mid-plane in the standard disk, 
the mean free paths are evaluated to be about 1cm and 500cm, respectively.  

Next, we describe motion of gas and grains.
The velocities of gas and grains, ${\boldsymbol{u}}$ and 
${\boldsymbol{U}}(m)$, 
are obtained in a similar way to Nakagawa et al.\ (1986),
though they considered equal-sized grains and
focused on the dust layer around the mid-plane. 
When grains have a size distribution,
the equations of motion for gas and grains are given by
\begin{eqnarray}
&&\frac{\partial {\boldsymbol{u}}}{\partial t}
 +({\boldsymbol{u}}\cdot \nabla){\boldsymbol{u}}
= -\int A\, m\, n(m) ({\boldsymbol{u}} - {\boldsymbol{U}}(m))\, dm
- \frac{G M_{*}}{|\boldsymbol{x}|^3} {\boldsymbol{x}}
- \frac{1}{\rho\sub{g}} \nabla P\sub{g}, 
\label{eq:motion_gas}
\\
&&\frac{\partial {\boldsymbol{U}}(m)}{\partial t}
+({\boldsymbol{U}}(m)\cdot \nabla){\boldsymbol{U}}(m)
= -A \rho\sub{g} ({\boldsymbol{U}}(m) - {\boldsymbol{u}}) 
- \frac{G M_{*}}{|\boldsymbol{x}|^3} {\boldsymbol{x}}. 
\label{eq:motion_dust}
\end{eqnarray}
The gas pressure $P\sub{g}$ is given by $c_s^2 \rho\sub{g}$.
At the surface layer directly irradiated by the central star,
grains suffers a strong radiative pressure (Takeuchi \& Lin 2003).
However, we neglected the radiation pressure in equation~(\ref{eq:motion_dust})
for simplicity.

We introduce a differentially rotating coordinate system with 
the angular velocity
$\varOmega = (GM_*/|{\boldsymbol{x}}|^3)^{1/2}$$\simeq 
\varOmega\sub{K} [1-(3/4)(z^2/R^2)]$ 
to eliminate the gravity of the central star in the equations of motion.
Instead of ${\boldsymbol{u}}$ and ${\boldsymbol{U}}(m)$, 
we use the velocities,
$\boldsymbol{v}={\boldsymbol{u}} - R\varOmega{\boldsymbol{e}}_\phi$ and 
$\boldsymbol{V}(m)={\boldsymbol{U}}(m) - R\varOmega{\boldsymbol{e}}_\phi$,
in the rotational frame. Since $\boldsymbol{v}$ and $\boldsymbol{V}(m)$ are 
much smaller than the Keplerian velocity, we can neglect 
second-order terms of them in the equations of motion. 
Then, setting $\partial/\partial t = \partial/\partial \phi = 0$,
we can rewrite equations (\ref{eq:motion_gas}) and (\ref{eq:motion_dust}) 
in the rotating cylindrical coordinates as (cf. Nakagawa et al. 1986)
\begin{eqnarray}
&&-\int A\, m\, n(m) (v_R-V_R)\, dm + 2\varOmega\sub{K} V_\phi 
- \frac{1}{\rho\sub{g}} \frac{ \partial P\sub{g} }{\partial R} = 0,
\label{eq:v_r1} \\
&&-\int A\, m\,  n(m) (v_\phi-V_\phi)\, dm - \frac12\varOmega\sub{K} v_R = 0,
\label{eq:v_o1} \\
&&-\int A\,  m\, n(m) (v_z-V_z)\, dm - \varOmega\sub{K}^2 z 
- \frac{1}{\rho\sub{g}} \frac{ \partial P\sub{g} }{\partial z} = 0,
\label{eq:v_z1} 
\end{eqnarray}
and
\begin{eqnarray}
&&-A\rho\sub{g}(V_R-v_R) + 2\varOmega\sub{K} V_\phi = 0,
\label{eq:V_r1} \\
&&-A\rho\sub{g}(V_\phi-v_\phi) - \frac12\varOmega\sub{K} V_R = 0,
\label{eq:V_o1} \\
&&-A\rho\sub{g}(V_z-v_z) - \varOmega\sub{K}^2 z = 0,
\label{eq:V_z1} 
\end{eqnarray}
respectively.
From equations (\ref{eq:V_r1}) and (\ref{eq:V_o1}),
the $R$- and $\phi$-components of the grain velocity are obtained as
\begin{eqnarray}
& & V_R(m) = 
\frac{\varGamma^2 v_R + 2\varGamma v_{\phi}}{1 + \varGamma^2}, 
\label{eq:V_r2} \\
& & V_{\phi}(m) = 
\frac{- \varGamma v_R + 2 \varGamma^2 v_{\phi}}
{2(1 + \varGamma^2)}, 
\label{eq:V_o2}
\end{eqnarray}
where $\varGamma = {\rho\sub{g} A}/{\varOmega\sub{K}}$.
Substituting equations (\ref{eq:V_r2}) and (\ref{eq:V_o2}) into 
(\ref{eq:v_r1}) and (\ref{eq:v_o1}), we also have for
the $R$- and $\phi$-components of the gas velocity
\begin{eqnarray}
& & v_R =
\frac{2 \lambda_1 \eta
R\varOmega\sub{K}}
{(1+\lambda_2)^2+\lambda_1^2},
\label{eq:v_r2} \\
& & v_{\phi} =
- \frac{(1+\lambda_2) \eta
R\varOmega\sub{K}}
{(1+\lambda_2)^2+\lambda_1^2},
\label{eq:v_o2}
\end{eqnarray}
where
\begin{equation}
\lambda_k = \int \frac{\varGamma^k}{1 + \varGamma^2}
\frac{m n(m)}{\rho\sub{g}}dm,
\qquad
\eta = - \frac{1}{2R\varOmega\sub{K}^2 \rho\sub{g}}
         \frac{\partial P\sub{g}}{\partial R}.
\label{eq:eta}
\end{equation}
Note that $\eta$ is dependent on $z$ as well as $R$. It should be also noted
that our definition of $\eta$ is different from that in Takeuchi \& Lin~(2002)
by the factor 2 but the same as Nakagawa et al.~(1986). 
The vertical velocity of the gas, $v_z$, is always negligibly small compared
with $V_Z$ as shown by Nakagawa et al.~ (1986). Hence, from equation~(\ref{eq:V_z1}) 
we have
\begin{eqnarray}
& & V_z = 
-\frac{z \varOmega\sub{K}^2}{\rho\sub{g}A},
\label{eq:V_z2} \\
& & v_z = 0.
\end{eqnarray}
The obtained velocities agree with those in Nakagawa et al.~(1986) 
in the case of the dust layer (i.e., $z/h_p\ll 1$) with equal-sized grains.
We use equations~(\ref{eq:V_r2})-(\ref{eq:V_z2}) at the calculation of the 
grain velocity in the simulations of dust growth and settling.

We finally explain the relative velocity between grains appearing
in the kernel $K_{m,m'}$. To evaluate the relative velocity between 
$\mu$m-sized grains, we have to take account of Brownian 
motion. The mean relative velocity due to Brownian motion is given by
(Suttner \& Yorke 2001) 
\begin{equation}
V\sub{B}(m,m') = \sqrt{\frac{8 k\sub{B} T}{\pi} \frac{m + m'}{m m'}}.
\label{eq:v_B}
\end{equation}
Then the total relative velocity between grains is defined by
\begin{equation}
\Delta V_{m,m'} = 
\sqrt{\left(\boldsymbol{V}(m) - \boldsymbol{V}(m')\right)^2 
+ V\sub{B}(m,m')^2}.
\label{eq:v_rel}
\end{equation}
%

\subsection{Numerical method}\label{subsec:method}

We calculated evolution of the grain distribution due to growth and settling,
by solving equation~(\ref{eq:coagsedi1})
with the grain velocities in the subsection 2.3. 
Because we neglected the radial drift of grains,
equation~(\ref{eq:coagsedi1}) can be integrated at each radial position, 
independently.
In our simulation, the disk extending from 0.1 AU to 150 AU
is logarithmically divided into 21 ring regions.
The median radius of the $i$-th ring region is given by 
$R_i = 	2^{(i-7)/2}$ AU and its width is defined by
$(2^{1/4}-2^{-1/4})R_i \simeq 0.35R_i$.   

Equation (\ref{eq:coagsedi1}) is integrated with respect to time with the 
first order accuracy. 
The change in the number density $n(m, z)$ 
is given by the sum of two changes.
One is the change due to coagulation,
which comes from two terms on the right-hand side of 
equation~(\ref{eq:coagsedi1}),
and the other is the change due to settling, 
which comes from the second term on the left-hand side.
These two changes are calculated separately at each time step.
The former change is calculated with the fixed bin scheme
(Nakagawa et al.\ 1981; Ohtsuki et al.\ 1990).
This scheme is valid for calculations of dust growth 
because runaway growth does not occur (Tanaka \& Nakazawa 1994).
The latter change is due to the advection term.
To calculate it accurately and stably, we used the spatially third-order MUSCL-type 
(Monotonic Upwind Scheme for Conservation Law-type) scheme developed by 
van Leer (1977).

As an initial condition for numerical calculations, all grains have
the radius of $0.1 \mu$m initially.
We assume the dust-to-gas ratio is uniform initially.
The density and the temperature of the gaseous disk 
are not varied with time during numerical simulations for simplicity. 

In the numerical integration of equation~(\ref{eq:coagsedi1}), 
the $z$-axis from 
$z = 0$ to $z = 2.5\sqrt{2} h_p$ is divided into 251 equally spaced grids
at each radial position. 
As for the coordinate of the grain mass,
we divide it into 500 discrete fixed mass bins.
The minimum mass bin is that of grains with the radius of $0.1 \mu$m
and its representative mass, $m_1$, is $4\pi \rho\sub{s} (0.1\mu\mbox{m})^3/3$.
The representative masses, $m_k$, of the smallest six bins are set to be
$km_1$ ($k\leq 6)$.
The larger mass bins are spaced logarithmically 
with a multiplication factor 1.15 (i.e., $m_k=1.15 m_{k-1}$).

\begin{figure}
\epsscale{0.8}
\plotone{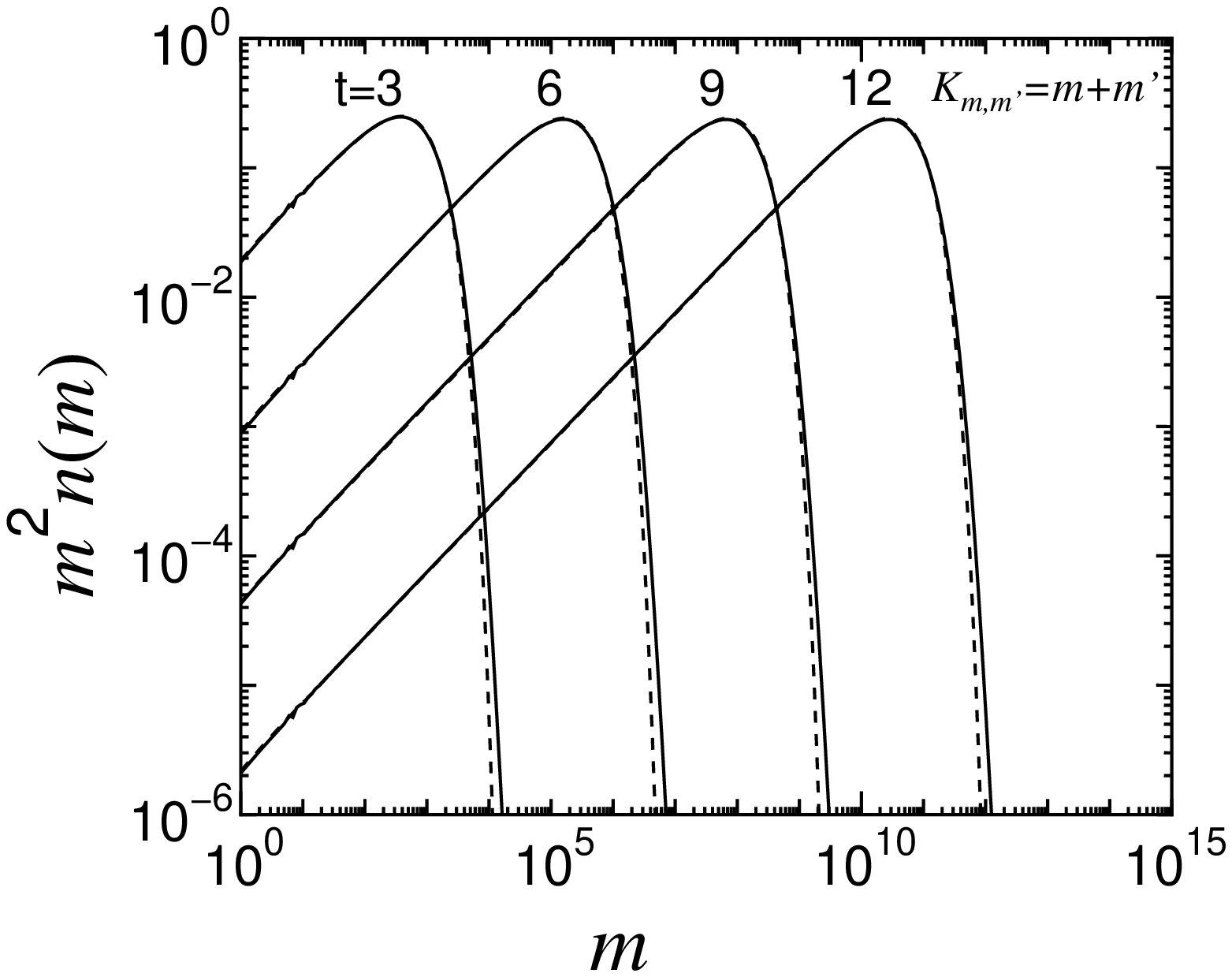}
\figcaption{Test of the coagulation part of our algorithm.
Numerical results (solid lines) are compared with analytic solutions to 
the coagulation equation (broken lines) for the kernel $K_{m, m'}=m+m'$.
\label{fig:1} }
\end{figure}

Time steps $\Delta t$ are determined as follows.
As the Courant condition, we made $\Delta t$ smaller than or equal to
$0.1 \Delta z/|v_{z,kj}|$, where $\Delta z$ is the space between vertical grids
and $v_{z,kj}$ is the settling velocity of grains in the $k$-th mass bin
of the $j$-th vertical grid.
We further restricted $\Delta t$ to solve the grain growth correctly.
Let $\Delta n_{kj}\super{coag}$ denote the change in the number density due to
coagulation at the mass bin of $m_k$ in the $i$-th vertical grid at a time
step.
The change $\Delta n_{kj}\super{coag}$ is proportional to $\Delta t$ in our 
scheme. We regulated the time interval $\Delta t$ so that the ratio 
$|\Delta n_{kj}\super{coag}|/n(m_k, z_j)$ is kept smaller than 0.025 
at all mass bins if $\Delta n_{kj}\super{coag}$ is negative.
Because of this condition, however, $\Delta t$
tends to be very small by mass bins that contain
negligibly small amount of grains.
To avoid it, if the mass contained in a mass bin
is smaller than $10^{-25}$ times the total mass of the simulation,
we removed all grains from the mass bin
at each time step to prevent it from regulating $\Delta t$.
Such removal of grains hardly breaks the mass conservation.
In our simulation with double precision, the relative error of 
the mass conservation is the order of $10^{-15}$. 
Furthermore, we checked the coagulation part of our algorithm,
by comparing numerical results with analytic solutions to the coagulation 
equation for the kernel $K_{m, m'}=m+m'$, as done by Ohtsuki et al.~(1990). 
Initially, all particles have an unit mass and they have an unit 
number density in this test.
Figure~\ref{fig:1} shows the result of the comparison between numerical
results and analytic solutions .
Numerical results agrees with analytic solutions much better than 
those of Ohtsuki et al.~(1990) because we adopted a small mass multiplication 
factor of 1.15.

\section{RESULTS ON GRAIN GROWTH AND SETTLING}\label{sec:result}
\subsection{Case at 8AU in the Standard Disk}\label{subsec:st8AU}

Before showing the numerical results, we mention
the growth time and the settling time, which have already been examined
in the previous studies (e.g., Safronov 1969; Weidenschilling 1980; 
Nakagawa et al. 1981, 1986).
According to Nakagawa et al. (1981), 
the growth time of grains, $t\sub{grow}$, is given by
\begin{equation}
t\sub{grow} = \frac{a}{\dot{a}} \simeq \frac{3}{\bar{n} p\sub{s}\sigma_{m,m}v}
 \simeq \sqrt{\frac{8}{\pi}}\frac{1}{p\sub{s}\zeta\sub{d} \varOmega\sub{K}}.
\label{eq:t_g}
\end{equation}
At the last equation, the grain number density $\bar{n}$ is estimated to be 
$\zeta\sub{d}\rho\sub{g}/m$; the collisional 
cross section $\sigma_{m,m}$ is given by $4\pi a^2$ under the assumption
of similar-size collisions\footnote{Our numerical results show that the 
dominant growth mode is coalescence 
between grains with similar sizes.}; and the settling velocity 
(eq.~[\ref{eq:V_z2}]) at $z=h_p$ is substituted to the relative velocity $v$. 
The growth time is independent of the grain size, the disk 
surface density, and the temperature. The settling time, $t\sub{settle}$, 
is given by
\begin{equation}
t\sub{settle} = \frac{z}{|V_z|} \simeq 
\frac{2\varSigma\sub{g}}{\pi e^{1/2}\rho\sub{s}\varOmega\sub{K}a},
\label{eq:t_settle}
\end{equation}
where we used equation~(\ref{eq:V_z2}) at the last equation, assuming $z=h_p$.
The settling time is also independent of the disk temperature and decreases 
with an increase in grain radius.
The initial grain size is so small (i.e., sub-micron size) that the growth 
time is much shorter than the settling time in an early stage.
Hence such grains grow considerably before settling down by a vertical 
distance comparable to $z$.
(Of course, since the grain growth is driven primarily by the settling 
velocity,
the settling is also essential to the growth of such small grains.)
As grains grow, the settling time eventually becomes comparable to 
the growth time. Then the grains do not grow much more before settling down to 
the dust layer (Nakagawa et al. 1986).  We can estimate such a critical grain 
radius, 
$a\sub{crit}$, by equating the settling time with the growth time.
From equations~(\ref{eq:t_g}) and (\ref{eq:t_settle}), we have
$a\sub{crit}= {p\sub{s}\varSigma\sub{d}}/{(\sqrt{2e}\rho\sub{s})}$,
where $\varSigma\sub{d}$ is the dust surface density.
The radius $a\sub{crit}$ corresponds to Safronov (1969)'s maximum radius to 
which grains can grow before settling to the dust layer.
Actually, $a\sub{crit}$ deviates only by a factor $\sim 3$
from the maximum radius obtained by Safronov (1969) with more exact treatment.

Now we show our numerical results on grain growth and settling.
Instead of the mass distribution
$n(m,z)$, we consider the distribution of the grain radius, $N(a,z)$, 
which is defined by $N(a,z) = n(m(a),z)({dm}/{da})$.  
Furthermore, by integrating $N(a,z)$ with respect to the height $z$, we 
define the surface number density of grains with each size, $N\sub{s}(a)$. That is,
\begin{equation}
N\sub{s}(a) = \int^\infty_{-\infty}N(a,z)dz.
\end{equation}
The surface number density $N\sub{s}(a)$ represents the vertically averaged 
size distribution.
The dimension of $N\sub{s}(a)$ is length$^{-3}$
because $N\sub{s}(a)da$ has the unit of number$/$length$^2$.
 
\begin{figure}
\epsscale{0.8}
\plotone{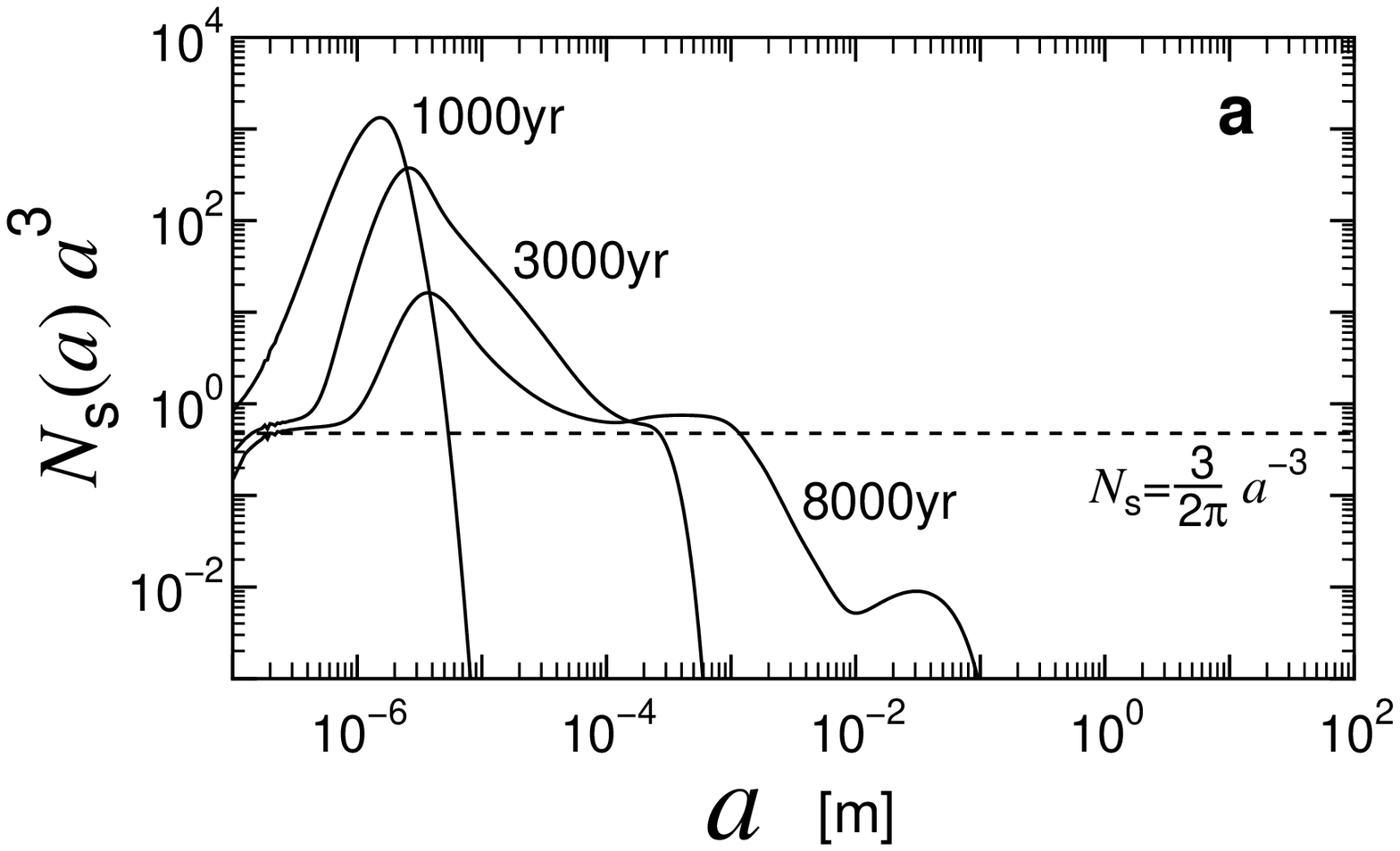}
\plotone{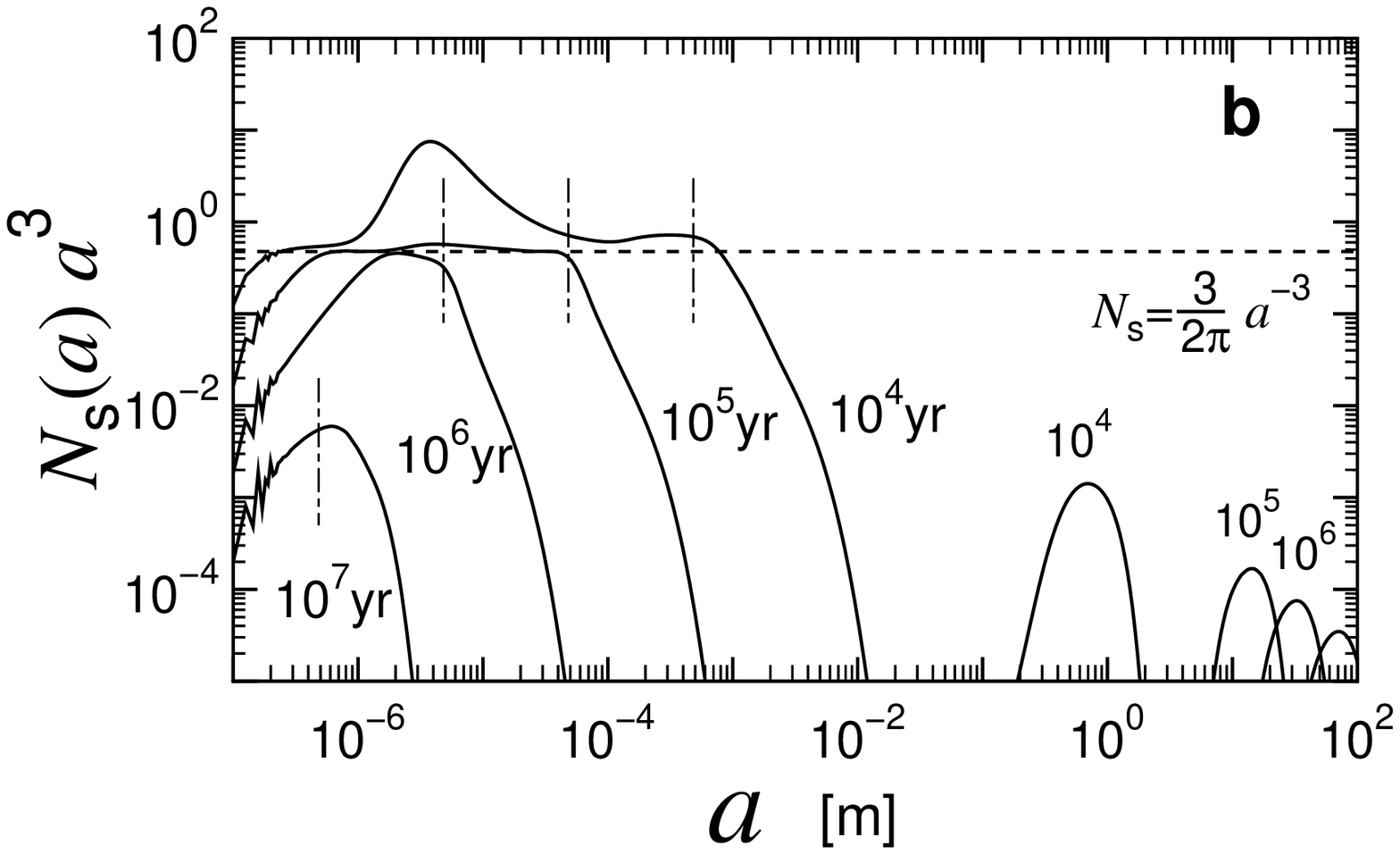}
\figcaption{Evolution of the vertically integrated size distribution 
$N\sub{s}(a)$ of grains at 8AU in the standard disk.
Panel~(a) shows the growth stage and panel~(b) corresponds to the 
settling stage. 
After formation of the dust layer ($t\ge 8000$yr),
the size distribution is divided into two parts:
floating small grains and large grains in the dust layer.
The size distribution of the floating grains is well-described
by the power-law distribution given by equation~(\ref{eq:size_dis})
(dashed lines).  Dashed and dotted lines in panel~(b) represent
the radii of the upper cut-off $a\sub{max}$ given by equation~(\ref{eq:a_max}).
Micron-sized grains remain floating above the dust layer until $10^6$yr.
\label{fig:2} }
\end{figure}

Figure~\ref{fig:2} shows evolution of the surface number density $N\sub{s}(a)$
at 8AU in the standard disk (\S 2.1), with $p\sub{s}=1$. 
Panel~(a) of the figure shows the size distributions in the 
early stage. Grains grow from 0.1$\mu$m-size to micron size in the first 
thousand years. At 3000 yr, the high-mass end of the distribution reaches 
several hundreds $\mu$m. From equation~(\ref{eq:t_g}), the growth time is
estimated to be 400yr at 8AU. This value is consistent with the growth
of  the high-mass end from 1000yr to 3000yr in Figure~\ref{fig:2}a. 
We found that the coalescence between grains with comparable masses
is dominant during this time interval in the simulation.
Thus the assumption of equal-sized collisions in equation~(\ref{eq:t_g})
would be valid.  At 8000yr, the size distribution has a bump 
at the high mass-end. Grains in the bump exist in the dust layer,
which formed near the mid-plane at that time. 
The formation time of the dust layer is 
roughly given by $\sim 10\, t\sub{grow}$, regardless of the disk model and
the radial position in disks. This result is consistent with results of the 
previous studies (e.g., Safronov 1969; Weidenschilling 1980; 
Nakagawa et al. 1986).

Evolution after the formation of the dust layer is shown in 
Figure~\ref{fig:2}b.
In this stage, the size distribution consists of two parts: 
large grains in the dust layer and small grains floating above the layer.
The floating grains gradually settle down onto the dust layer without
considerable growth.
Relatively large floating grains settle onto the layer earlier and 
the upper cut-off of the size distribution of floating grains shifts to the 
smaller size. At $t \sim 10^6$yr, only micron-sized grains remain floating. 
In this way, the time-evolution can be divided into two stages: the 
growth stage (Fig.~\ref{fig:2}a)\footnote{
It should be noted that the grain settling is essential even in the growth 
stage because the grain growth is driven primarily by the settling velocity.
At the end of the growth stage, the dust layer is formed through the grain 
settling.}
and the settling stage (Fig.~\ref{fig:2}b).
The dust layer is formed at the transition between the two stages

In the settling stage, the size distribution of floating grains is 
well-described by a power-law distribution of $N\sub{s}(a) \simeq 0.5 a^{-3}$.
Dashed lines in Figure~\ref{fig:2} are the size distribution given by 
equation~(\ref{eq:size_dis}), which will be derived in the subsection 3.3.
At $t=10^4$yr, the size distribution has a bump at $a\sim 4\mu$m.
Grains in this bump exist at $z<h_p$.
These grains are removed by the coalescence with larger floating grains 
settling faster in the early part of the settling stage.

The size distribution of floating grains is also characterized by
its upper cut-off. The grain radius at the upper cut-off, $a\sub{max}$, can
be analytically estimated. Since $a\sub{max}$ decreases due to grain 
settling, the radius $a\sub{max}(t)$ is estimated from 
$t\sub{settle}(a\sub{max}) \simeq t$. 
Introducing a numerical factor in this relation for fitting with 
the numerical results, we obtain 
\begin{equation}
a\sub{max}(t) =  2.5
\frac{\varSigma\sub{g}}{\rho\sub{s} \varOmega\sub{K} t}. 
\label{eq:a_max}
\end{equation}
In Figure~\ref{fig:2}b, $a\sub{max}$ of equation~(\ref{eq:a_max}) is plotted 
with vertical dotted and dashed lines at each time.
The upper cut-off radii in the numerical results are well-described by 
equation~(\ref{eq:a_max}).

At the low-mass end, the surface number density decreases with time
even in the settling stage. This is due to grain growth by Brownian motion. 
As grains approach the mid-plane, their settling velocities decrease while 
Brownian motion do not vary. Thus, since the collision rate between grains 
is proportional to their relative velocity,
Brownian motion is effective on grain growth in the settling stage
especially for small grains.
Since smaller grains remain floating at a higher altitude,
growth due to Brownian motion affects the opacity at the disk surface.
Furthermore, growth due to Brownian motion is remarkable in an inner 
part of the disk rather than an outer part because of high temperature 
at the inner part.  
Brownian motion and settling deplete floating grains
at the low-mass and the high-mass ends, respectively.
These reduce the disk optical thickness in the settling stage.

In the dust layer, grains grow to more than 100-meter-size
at $t \geq 1 \times 10^5$ yr (see Fig.~\ref{fig:2}b). 
However, grain growth in the layer would not be calculated accurately
because of the following two problems.
The first problem is insufficient vertical resolution.
The vertical resolution $\delta z$, which is $\sim 0.01h\sub{p}$ in our 
calculation, gives the minimum thickness of the dust layer.
Because of this limit, the dust-gas ratio has an upper limit of the order of 
unity in our simulations. 
Nakagawa et al. (1986) investigated dust growth in the dust layer 
getting thinner infinitely due to dust settling
and showed that dust growth terminates when the gas-dust ratio becomes much 
larger than unity.  
Such termination of dust growth does not occur in our simulations
since the coarse resolution keep the dust-gas ratio $\lesssim 1$.
However, the infinitely shrinking dust layer assumed by
Nakagawa et al. (1986) would be unrealistic.
Many authors reported that turbulence due to the shear instability
prevents the further dust settling when the gas-dust ratio exceeds unity
(e.g., Weidenschilling 1980; Cuzzi et al. 1993).
Hence termination of dust growth proposed by Nakagawa et al. (1986) would not
occur in the realistic disk. The second and more serious problem is 
neglect of the radial drift of grains.
If the radial drift is taken into account, particles with meter-sized
would rapidly fall to the central star and be lost (e.g., Adachi et al. 1976).
In spite of above problems in our calculation, 
the inaccuracy in the dust layer would not significantly influence
our results on the disk optical thickness
because the dust layer has only a minor contribution to the disk optical 
thickness in most cases. 

\begin{figure}
\epsscale{0.7}
\plotone{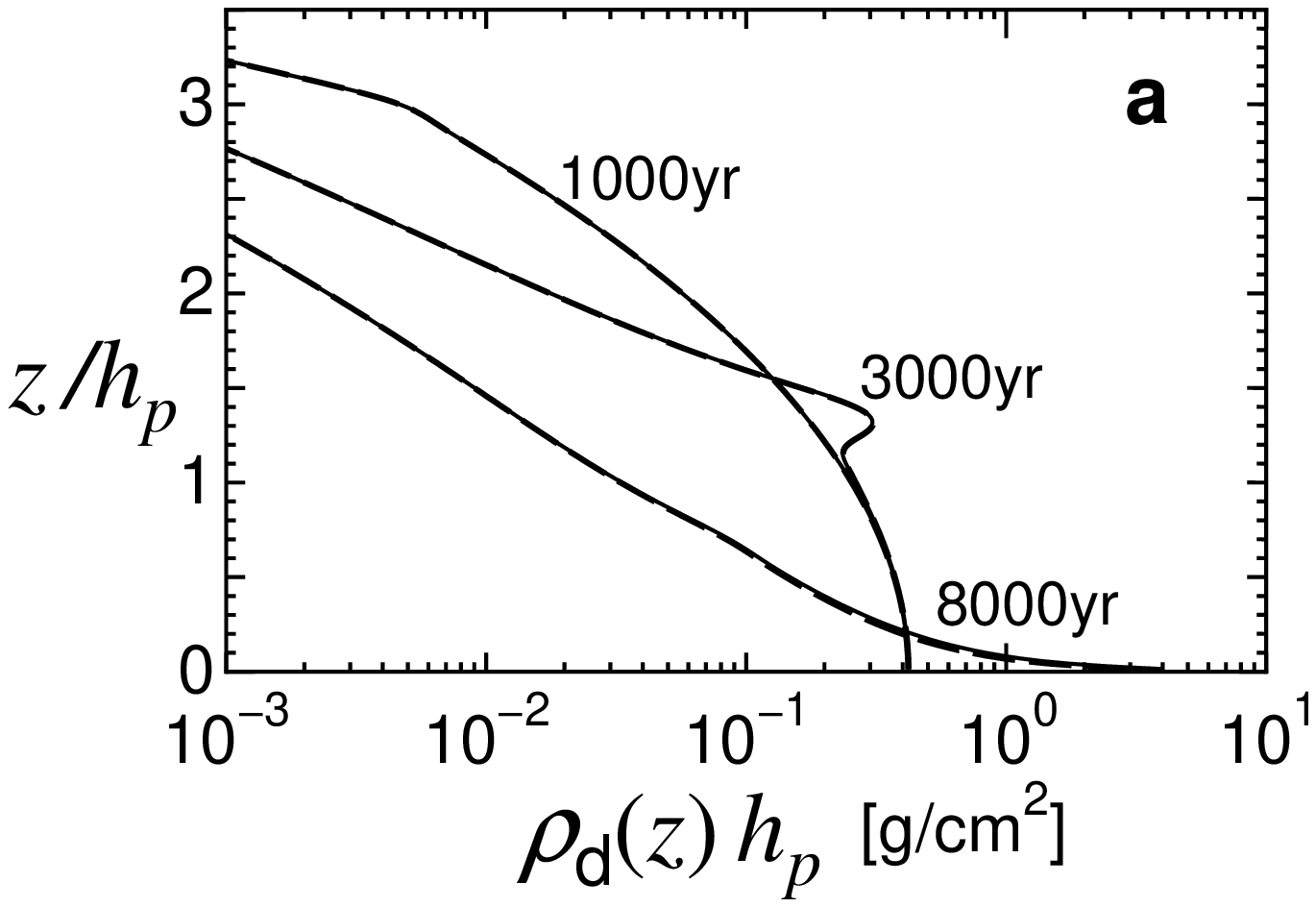}\vspace{3mm}
\plotone{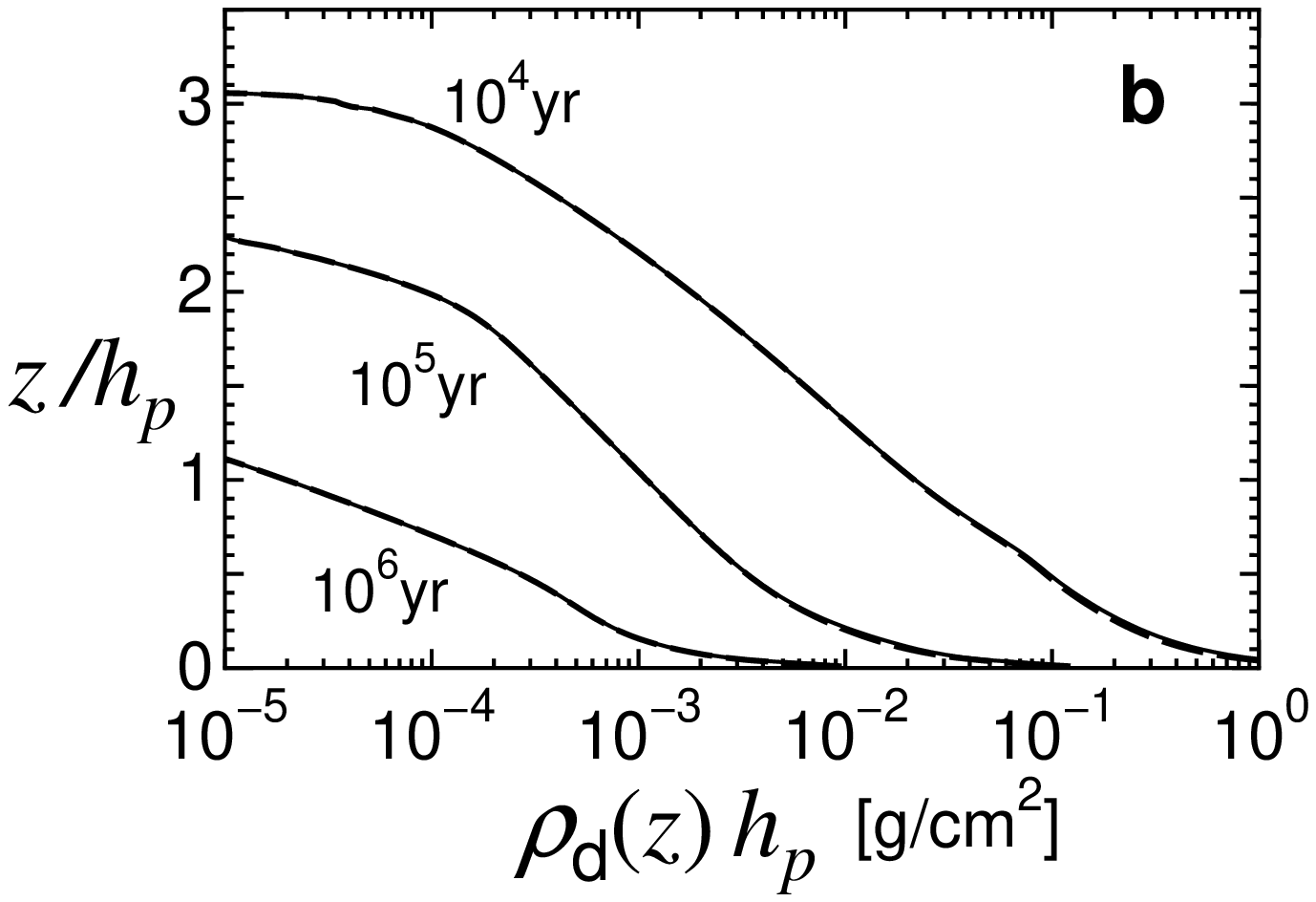}
\figcaption{Evolution of the vertical density distribution 
$\rho\sub{d}(z)$ of grains at 8AU in the standard disk.
Because of settling, the grains deplete with time especially at
a high altitude of $z \ge h_p$.
We also plotted the density distributions in a disk with 
temperature three times as high as the standard disk with dashed lines.
Both cases almost agree with each other.
\label{fig:3} }
\end{figure}

In Figure~\ref{fig:3}, vertical density distributions of dust grains are shown.
The vertical axis in the figure is $z$ and the horizontal axis is the 
density of grains, $\rho\sub{d}$, given by $\int m(a) N(a,z)da$. Panel~(a) 
shows evolution in the growth stage and Panel~(b) corresponds to the 
settling stage. At 8000 yr, the density of grains is significantly
enhanced near the mid-plane and the dust layer forms there.
Grains are depleted with time especially at a high altitude of $z \ge h$.

\subsection{Parameter Dependences}\label{subsec:para}
The dust growth and settling are expected to be dependent on the following 
physical parameters: the disk surface densities of gas and dust grains, 
the temperature, the material density of grains, the sticking probability, 
and the radial position in the disk (i.e., $R$). 
Here we examine the dependence of the size distributions of floating grains 
on these physical parameters.

\begin{figure}
\epsscale{0.6}
\plotone{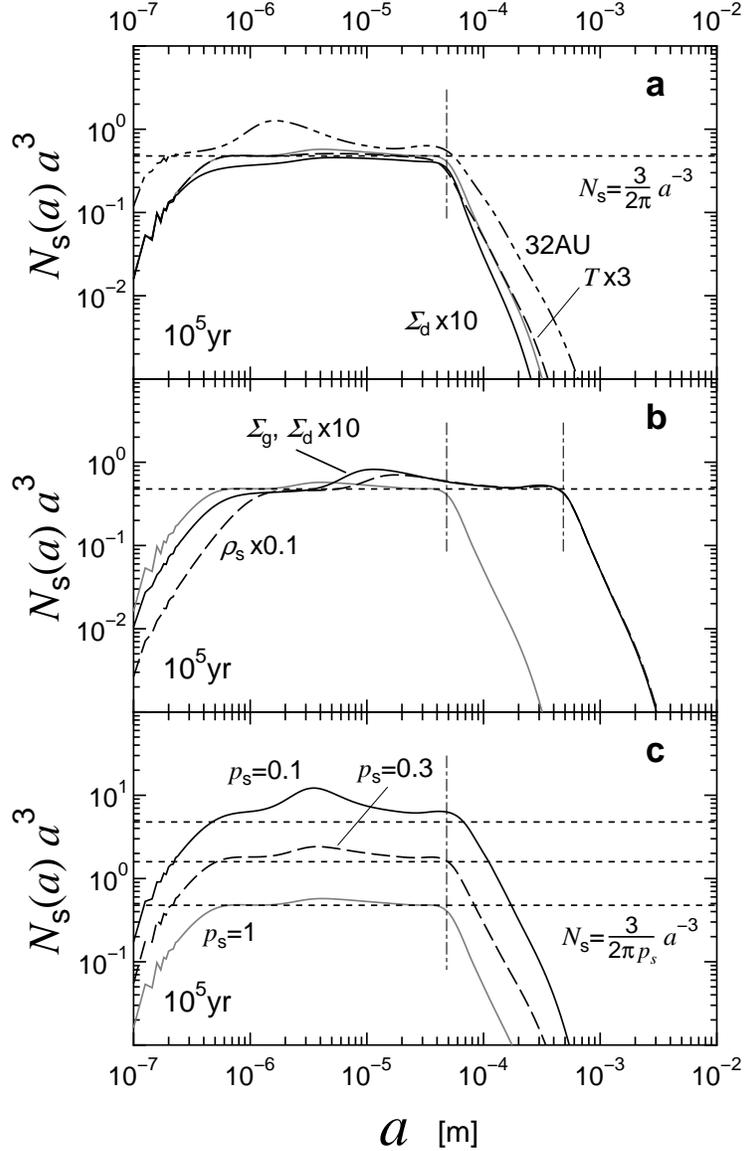}
\figcaption{Parameter dependence of the size distribution
of floating grains at $t=10^5$yr.
The size distribution at 8AU in the standard disk is
plotted by gray lines at each panel for comparison.
Panel~(a) displays a case at 32AU (the dashed and dotted line) and 
cases with a high dust-to-gas ratio (the solid line)
and with a high temperature (the long dashed line).
Panel~(b) shows cases with a high disk surface density (the solid line)
and with a low material density of grains (the long dashed line).
Panel~(c) shows the dependence of the size distribution on the
sticking probability $p\sub{s}$.
In all cases, the size distribution and its upper cut-off
are well-described by equation~(\ref{eq:size_dis}) (dashed lines)
and equation~(\ref{eq:a_max}) (dashed dotted lines), respectively.
\label{fig:4} }
\end{figure}

In Figure~\ref{fig:4}, we show size distributions with different values of the 
physical parameters. First we examine the dependence on the surface density 
of dust ($\varSigma\sub{d}$), the disk temperature, and the radial position. 
To do this, 
another simulation similar to the above was done at 32AU in the standard disk.
Furthermore, two more runs were done at 8 AU in disks with the dust surface 
density (or equivalently the dust-to-gas ratio $\zeta\sub{d}$) ten times as 
large as that of 
the standard disk and with the temperature three times higher.
In Figure~\ref{fig:4}a, these three cases were compared with the standard case in 
Figure~\ref{fig:2} at $t=10^5$yr. These all cases have the same size distribution for 
floating grains in the settling stage. The upper cut-off radii of the size 
distributions also have the same value. This is because $\zeta\sub{d}$ and 
$T$ are not included in equation~(\ref{eq:a_max}) and 
$\varSigma\sub{g}/\varOmega\sub{K}$ is independent of $R$ in the standard disk.
We also mention distributions in the growth stage. In the growth stage, 
the case at 32AU and the case with a high dust-to-gas ratio have different 
size distributions from others because of their different growth time. 
On the other hand, the difference in the disk temperature hardly 
affects the grain size distribution in the growth stage, too.

In Figure~\ref{fig:4}b, we showed the dependence on the gas surface density
and the material density of grains.
We did runs with a low material density of $\rho\sub{s}=0.14$gcm$^{-3}$ 
and with the surface densities of both gas and grains ten 
times as large as those of the standard disk.
These runs also result in the same power-law size distribution.
The upper cut-off radius in these cases are larger than
that of the standard case but they are also well-described with 
equation~(\ref{eq:a_max}).

Finally we examine the dependence on the sticking probability. 
Performing two more runs with $p\sub{s}=0.1$ and 0.3,
we compared these results with the standard case in Figure~\ref{fig:4}c.
The power-law distributions form even for different sticking 
probabilities but the absolute values of the distribution
are inversely proportional to the sticking probability.

In all simulations, the mass of the central star was set to be one
solar mass. The central star's mass is included only in $\varOmega\sub{K}$
in the equations governing the simulations.
If the time and velocities are scaled by $\varOmega\sub{K}^{-1}$ and
$R\varOmega\sub{K}$,respectively,
$\varOmega\sub{K}$ appears only in the ratio $h_p/R$.
In the above, we found that the numerical results does not depend
on $h_p/R$ (or the disk temperature). 
Hence, the above numerical results can be used for
stars with different masses if we use the scaled time. 

In the above we showed that, in the settling stage,
the size distribution of floating grains is universally given by
a power-law distribution with the index of $-3$,
that is, $N\sub{s}=C a^{-3}$.
The coefficient $C$ of the distribution is inversely
proportional to the sticking probability and independent of
the disk properties and the material density of grains.
The upper cut-off of the size distribution is well-described by
equation~(\ref{eq:a_max}).

\subsection{Analytical Solution of the Grain Size Distribution}
\label{subsec:analytic}
	
\begin{figure}
\epsscale{0.7}
\plotone{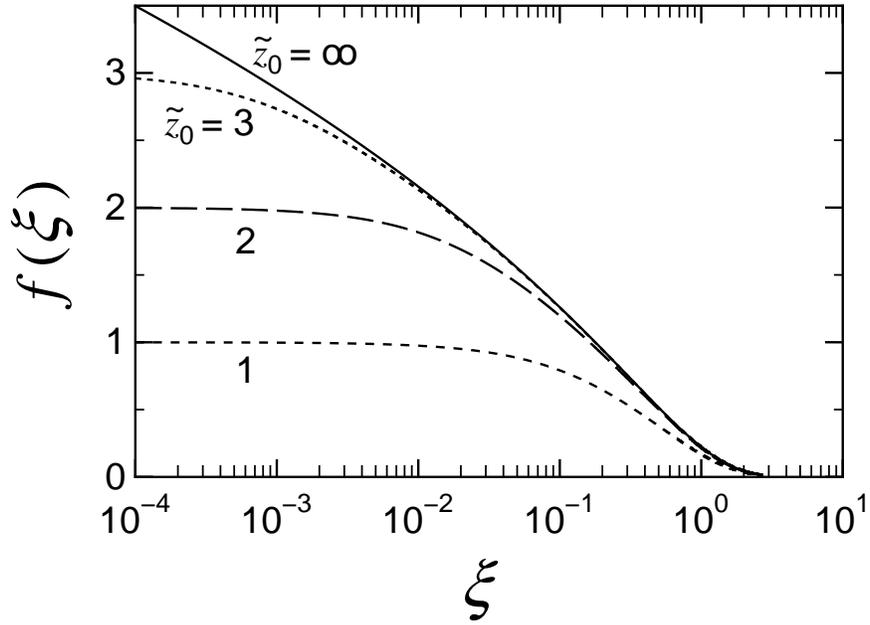}
\figcaption{Settling trajectories of grains without growth in isothermal 
protoplanetary disks for various initial altitudes $\tilde{z}_0$.
\label{fig:5} }
\end{figure}

Here we analytically derive the grain size distribution in the settling stage.
First we examine the settling process.  Here we neglect grain growth since 
it hardly occurs for floating grains in the settling stage.
The settling velocity is given by 
equation~(\ref{eq:V_z2}). 
We consider Epstein's regime for gas drag
in equation~(\ref{eq:gas_drag4}) because most floating grains are 
smaller than cm-size in the settling stage. The vertical distribution of 
the gas density is generally given by 
\begin{equation}
\rho\sub{g}(z)=\frac{\varSigma\sub{g}\varOmega\sub{K}}{
c_s} g(z/h_p).
\label{eq:rho_gas2}
\end{equation}
This expression is valid even for the cases of vertically non-isothermal disks.
In non-isothermal cases, we set $h_p$ to be the value at the mid-plane to make
it independent of $z$. In the isothermal case, the function $g$ is given by 
$\exp[-z^2/(2h_p^2)] /\sqrt{2\pi}$. The grain settling is described by the 
equation
\begin{equation}
\frac{dz}{dt} 
= - \frac{ c_s \varOmega\sub{K} z}{\varSigma\sub{g} A(a) g(z/h_p)}.
\label{eq:settle1}
\end{equation}
In Epstein's regime, the drag coefficient $A(a)$ is inversely proportional to
$a$ and the ratio $A/c_s$ is independent of $z$ even for non-isothermal disks.
We introduce the normalized vertical axis $\tilde{z}=z/h_p$ and 
the non-dimensional time 
$\xi = c_s \varOmega\sub{K} t/ (A\varSigma\sub{g})$.
Then equation~(\ref{eq:settle1}) is reduced to
$d\tilde{z}/d\xi = - \tilde{z}/g(\tilde{z})$ and the solution is 
expressed as $\tilde{z}=f(\xi)$. That is,
\begin{equation}
z=h_p \, f\!  
\left(
\sqrt{\frac{\pi}{8}}\frac{\rho\sub{s}a\varOmega\sub{K} t}{\varSigma\sub{g}}
\right).
\label{eq:zsol}
\end{equation}
The above solution also depends on the initial height ${z}_0$.
Figure~\ref{fig:5} shows the solution $f(\xi)$ for various $z_0$
in the isothermal case.
If the initial height is high enough ($\tilde{z}_0 >1$), the solution is almost
independent of $\tilde{z}_0$ except at an early stage. This is
because the gas density is very low at a high altitude and grains settle
down rapidly in Epstein's regime. 
The small floating grains in Figure~\ref{fig:2}b come from the high altitude.
On the other hand,
most grains initially located at a low altitude ($\tilde{z}_0 \lesssim 1$)
are removed by the coagulation with larger ones settling from a high altitude.
Thus we only consider grains initially located at a high altitude and 
neglect the dependence of the solution on ${z}_0$.
Equation~(\ref{eq:zsol}) represents well-defined trajectory of each grain.
This indicates that there exist grains with a definite size at each height 
at certain time. 
That is, the settling causes size segregation in the disk.
Actually, in our numerical simulations, the altitude of each floating
grain is well-described by Equation~(\ref{eq:zsol}) and the size segregation
occurs.
This size segregation makes the relative velocity between colliding grains
small and prevents grain growth considerably in the settling stage.

Secondly, we derive a simple relation between the
spatial and size distribution $N(a,z)$ and the vertically integrated
size distribution $N\sub{s}(a)$.
Because of the size segregation,
the spatial and size distribution $N(a',z)$ is proportional to
the delta function $\delta(a'-a(z))$, where the function $a(z)$
is given by solving equation~(\ref{eq:zsol}) with respect to $a$.
In a vertical section of ($z$, $z+\Delta z$), there exist
grains with sizes ($a(z)$, $a(z)+\Delta a$), where
$\Delta a$ is given by $\Delta z/(dz/da) = \Delta z a/(V_z t)$
(see eq.~[\ref{eq:zsol}]).
Noting that grains with a same size exist at two altitudes (i.e.,
above and below the mid-plane), we obtain
$2\Delta z \int da' N(a,z) = |\Delta a|N\sub{s}(a(z))$.
This equation determines $N(a',z)$ as
\begin{equation}
N(a',z) = \frac{1}{2} \delta(a'-a(z)) \frac{a(z)N\sub{s}(a(z))}{|V_z|t}.
\label{eq:n-ns}
\end{equation}
%

Using equation~(\ref{eq:n-ns}), we can derive the expression of 
the power-law size distribution of $N\sub{s}(a)$ in the settling stage.
In the growth stage, the growth time $t\sub{grow}$ is shorter than 
$t\sub{settle}$. Dust settling, however, decreases the grain number density 
at each altitude and increases the growth time. 
At the transition to the settling stage, the two time scales equal to each 
other and, in the settling stage, the growth time further increases
because of the size segregation. Hence the size distribution is fixed at the 
transition (see also Fig.~\ref{fig:2}) and the fixed size distribution is 
obtained from the equation of $t\sub{grow}=t\sub{settle}$.
We reevaluate $t\sub{grow}$, by setting the grain number density 
$\bar{n}(z) = \int N(a',z)da'$ in equation~(\ref{eq:t_g})
and using equation~(\ref{eq:n-ns}) for $N(a',z)$.
The time $t$ can be replaced by $t\sub{settle}$
since equation~(\ref{eq:n-ns}) is derived from the solution 
of settling. Then we obtain
\begin{equation}
t\sub{grow} = \frac{ 3t\sub{settle} }{2\pi p\sub{s} N\sub{s}(a)a^3} 
\end{equation}
and the equation, $t\sub{grow}=t\sub{settle}$, yields 
\begin{equation}
N\sub{s}(a) = \frac{3}{2\pi p\sub{s}a^3}.
\label{eq:size_dis}
\end{equation}
The obtained size distribution is dependent only on the sticking probability.
In Figures~\ref{fig:2} and \ref{fig:4}, equation~(\ref{eq:size_dis}) is 
plotted by dashed lines.
In all cases,  equation~(\ref{eq:size_dis}) agrees well with the numerically 
obtained size distributions of floating grains in the settling stage. 

In the above derivation, 
we only consider grains initially located at a high altitude ($\tilde{z}_0>1$)
to make the solution almost independent of $z_0$.
Thus grains initially located at $z\le h_p$ are not
included in the size distribution of equation (\ref{eq:size_dis}).
In Figure~\ref{fig:2}b, such grains are located in the bump at several $\mu$m
at 10$^4$yr.
Since most of them are removed by the coagulation with larger grains
in an early part of the settling stage,
equation (\ref{eq:size_dis}) is thus valid in most of the settling 
stage.\footnote{For Stokes' regime, we can obtain a similar
size distribution with almost same derivation.
In Stokes' regime, however, the solution of equation (\ref{eq:zsol})
is dependent of $z_0$ even for large $z_0$.
Thus the above derivation would be invalid in Stokes' regime.} 
Furthermore, since we did not assume vertically isothermal disks,
equation~(\ref{eq:size_dis}) would be valid for vertically 
non-isothermal disks like the two-layer model, too.

We also mention the temperature dependence of the spatial and size 
distribution $N(a,z)$. As seen from equations~(\ref{eq:n-ns}) and 
(\ref{eq:settle1}), $N(a,z)$ depends on the temperature.
However, if the spatial and size distribution is considered as a function
of $a$ and the normalized vertical axis $\tilde{z}$ (i.e., $N(a,\tilde{z})$),
the product $h_p N(a,\tilde{z})$ is independent of the temperature.
This invariance of $h_p N(a,\tilde{z})$ is also seen in numerical simulations.
In Figure~\ref{fig:3}, we also plotted the result of the high temperature case
to compare with the standard case.
In both of the growth stage and the settling stage, 
the disk temperature hardly affect the values of the product 
$h_p \rho\sub{d}(\tilde{z})$ $(=\int m(a) h_p N(a,\tilde{z})\, da)$.
This result comes from the invariance of $h_p N(a,\tilde{z})$.
In our simulation of dust growth and settling, we used a tentative disk 
temperature. Owing to the invariance of $h_p N(a,\tilde{z})$, however, 
our numerical results can be used for disks with different temperature 
distributions.
\section{EVOLUTION OF DISK SPECTRAL ENERGY DISTRIBUTIONS}
\label{sec:SED}

\subsection{Disk Optical Thickness}\label{subsec:thickness}

In order to evaluate the disk optical thickness,
we adopted the following grain model. Grain compositions 
and abundance were set to be the same as Pollack et al.~(1994).
That is, the compositions are H$_2$O-ice, organics, olivine, pyroxene, 
metallic iron, and troilite.
We assumed that grains consist of a mixture of these compositions
instead of the core-mantle type grains adopted in Chiang et al.~(2001). 
The model of well-mixed grains would be suitable for grown grains 
(Miyake \& Nakagawa 1993).
As the optical constants of olivine and pyroxene, we used data by
Dorschner et al. (1995).
For H$_2$O-ice, we used the same data as Miyake \& Nakagawa (1993).
For troilite, data by Begemann et al. (1994) was used
at wavelengths from 10$\mu$m to 500$\mu$m.
Other optical constants are set to be
the same as those in Pollack et al.~(1994).

For such grains, we calculated the grain emissivity (or the absorption 
efficiency) $\varepsilon_\nu(a)$ for each grain radius $a$, using the 
Mie theory (Bohren \& Huffman 1983). 
At high temperature of $T>160$K, we used the grain emissivity of grains 
without ice or organics while we used that including ice or organics
at $T<160$K. By the use of the emissivity 
$\varepsilon_\nu(a)$ and the spatial and size distribution $N(a, z)$, 
the opacity $\kappa_\nu$ is expressed by
\begin{equation} 
\kappa_\nu = \frac{1}{\rho\sub{g}}
\int \varepsilon_\nu(a) \pi a^2 N(a, z) da.
\label{eq:kappa}
\end{equation}
Since $\rho\sub{g} \propto h_p^{-1}$ (see eq.~[\ref{eq:dens}] or
[\ref{eq:rho_gas2}]), we can see that the opacity is expressed with the 
product $h_p N(a,\tilde{z})$, which is independent of the disk temperature.
The disk optical thickness $\tau_\nu$ is given by 
$\int^\infty_{-\infty} \rho\sub{g}\kappa_\nu dz$.

\begin{figure}
\epsscale{0.7}
\plotone{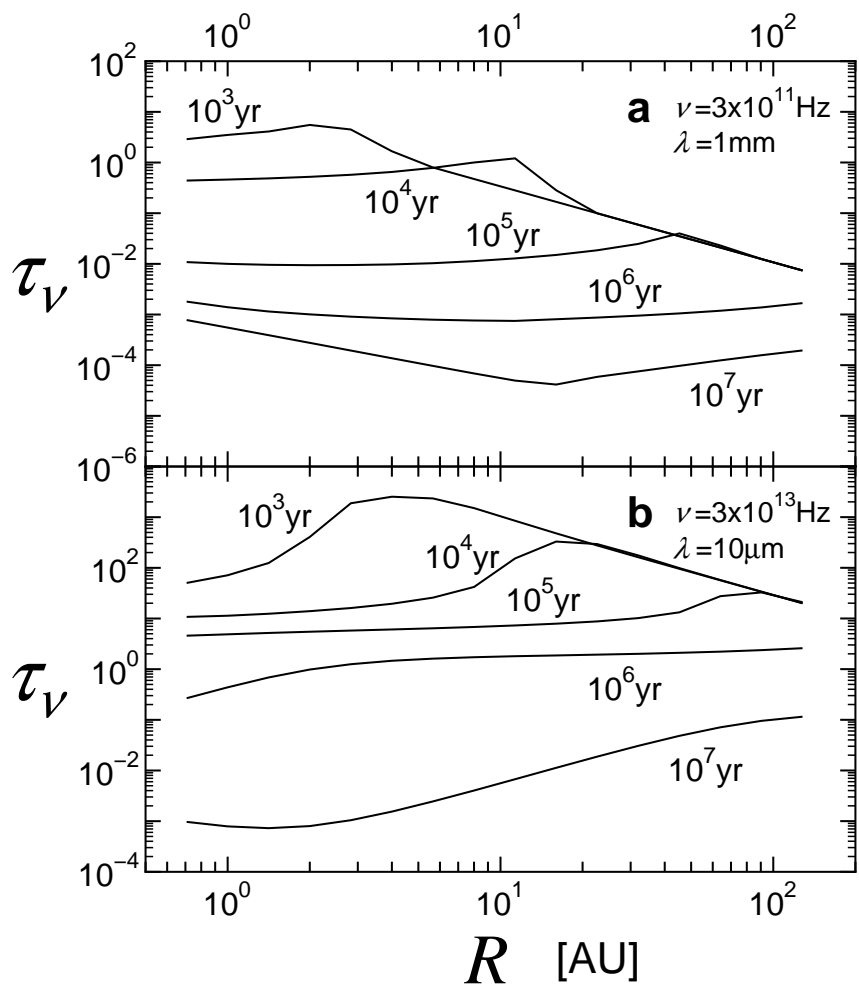}
\figcaption{Radial distributions of the disk optical thickness 
are displayed at each time for the standard disk.
Panel~(a) show the optical thickness for radiation with $\lambda=$ 
1mm and panel~(b) is that for $\lambda = 10\mu$m.
\label{fig:6} }
\end{figure}

At each heliocentric distance $R$ in the standard disk,
we performed simulations of dust growth and settling and calculated the 
disk optical thickness. 
Figure~\ref{fig:6} shows the radial distributions of the optical thickness in 
the standard disk for wavelengths $\lambda = 10\mu$m and 1mm. 
From $10^4$ to $10^6$yr, 
the radial distribution of the optical thickness is almost 
flat for both wavelengths.
This is because the vertically integrated size distribution $N\sub{s}(a)$ of 
floating grains and its upper cut-off $a\sub{max}$ are independent of the 
radial position $R$.
Large grains in the dust layer have only a minor contribution
to the optical thickness though they are effective in the inner part
($<$10AU) for $\lambda =1$mm at $t=10^7$yr.
For $\lambda =10\mu$m, the optical thickness is
much smaller in the inner disk than the outer disk.
This is because grain growth due to Brownian motion
depletes the micron-sized grains in the inner disk
(see Fig.~\ref{fig:2}b).
This extremely optically thin region may corresponds to
inner holes recently observed by the Spitzer Space Telescope 
(Forrest at al. 2004; Meyer et al. 2004; Uchida et al. 2004). 
In most previous analysis of observations, the optical thickness is assumed to 
be proportional to the disk surface density.  Under this assumption, 
the radial power-law index of the surface density $\varSigma\sub{g}(R)$ 
is estimated to be from 0 to -1 by recent observations with high-resolution at
millimeter wavelengths (Kitamura et al. 2002).
However, this assumption is invalid because of grain growth.
Our model shows that the flat radial distribution of the disk optical 
thickness forms even in disks in which the surface density is proportional 
to $R^{-3/2}$ like the MMSN.

\begin{figure}
\epsscale{0.7}
\plotone{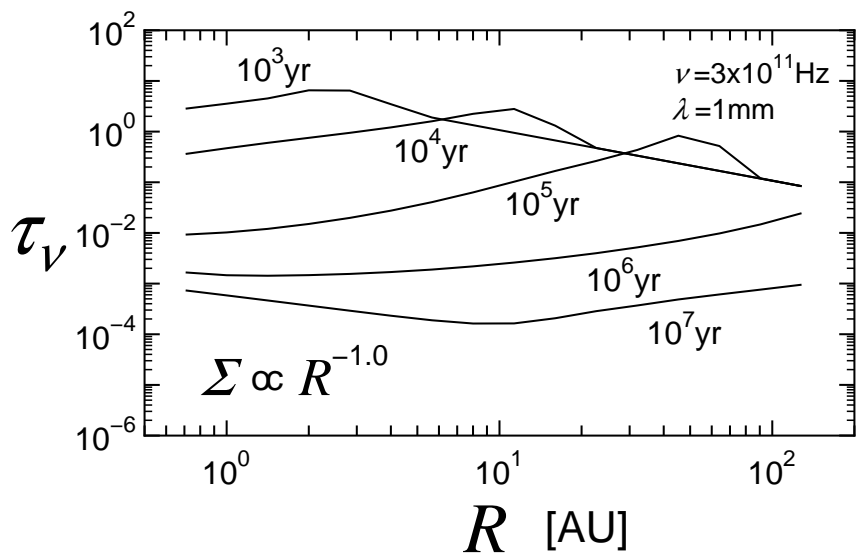}
\figcaption{Same as Fig.~\ref{fig:6}a but in a disk in which the surface 
density is proportional to $R^{-1}$.
\label{fig:7} }
\end{figure}

We also performed simulations for another disk, in which the surface 
density is proportional to $R^{-1}$. The gas surface density at 1AU, 
$\varSigma\sub{g1}$, is set to be the same as the standard disk.
In Figure~\ref{fig:7}, we display the radial distributions of the optical 
thickness of 
this disk for $\lambda = 1$mm. The optical thickness is almost proportional to 
$R^{1/2}$ at $t>10^4$yr in this disk. This is because the upper cut-off radius 
$a\sub{max}$ of the size distribution is also proportional to $R^{1/2}$ at 
a given time in this disk (see eq.~[\ref{eq:a_max}]). 
From these results, it is found that the 
difference in the power-law index between $\tau_\nu$ at millimeter 
wavelength and $\varSigma\sub{g}$ is about 1.5 at $t \lesssim 10^6$yr.

\subsection{Disk Structure and Spectral Energy Distributions}
\label{subsec:seds}

Using our results on grains and the two-layer model proposed by 
Chiang \& Goldreich (1997), we calculate disk temperatures and 
spectral energy distributions of disks.
In the two layer model, the disk consists of two surface layers 
(i.e., the upper and the lower) and the disk interior.
Then the disk structure is characterized by
the heights of the surface layers, $\pm z\sub{s}(R)$, and temperatures of
the surface layers and the disk interior, $T\sub{s}(R)$, $T\sub{i}(R)$.
We neglected the effect of the inner rim where silicate grains
evaporate because it has a only minor contribution to the SED
for T Tauri disks (Dullemond et al. 2001).
We also neglected heating due to the disk accretion
as Chiang \& Goldreich (1997) did.
Equations governing the two-layer model have already been
described in detail by Chiang et al. (2001) and, thus, we 
briefly describe them in Appendix A.

\begin{figure}
\epsscale{0.7}
\plotone{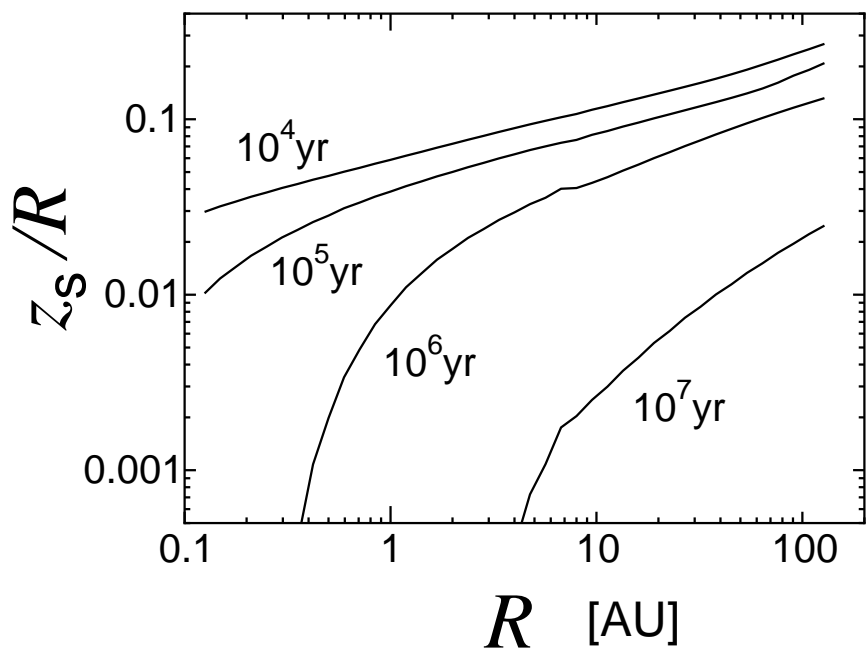}
\figcaption{Evolution of the height of the surface layer in the standard disk.
The surface layer falls because of grain settling and growth especially in 
an inner part of the disk.
\label{fig:8} }
\end{figure}

We numerically solved the equations and obtained the structure of the 
standard disk.
The effective temperature, $T_*$, and the radius, $R_*$ of the central star
are set to be 4000K and 0.0097AU, respectively.
Figure~\ref{fig:8} shows time-evolution of the height of the upper surface
layer $z\sub{s}$.
The surface layer descends because of grain settling and growth.
The descent of the surface layer is remarkable in an inner part of the disk.
This is because of the significant reduction of 
the optical thickness in the inner disk mentioned above.
Actually, the disk is optically thin to the starlight in the region of 
$R<0.25$AU ($R<3.4$AU) at $10^6$yr ($10^7$yr) and the disk interior 
disappears there.
The grazing angle between the starlight and the disk surface
given by equation~(\ref{eq:angle}) decreases because of the descent of the
disk surface. However, the grazing angle remains positive 
and the self-shadowing does not occur.

\begin{figure}
\epsscale{0.55}
\plotone{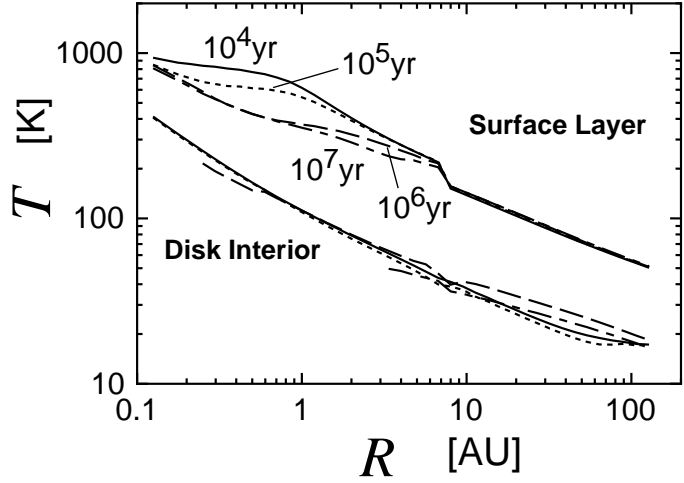}
\figcaption{Temperature distributions in the surface layer and
the disk interior of the standard disk at various time.
Solid lines, dotted lines, dotted and dashed lines, and long dashed lines
corresponds to $t= 10^4$, $10^5$, $10^6$, and $10^7$yr, respectively.
The temperature in the disk interior is much lower than that in the 
surface layer. The temperature in the disk interior does not vary much
with time.
\label{fig:9} }
\end{figure}

The temperatures in the surface layer and the disk interior
are shown in Figure~\ref{fig:9}.
Because the temperature of the surface layer depends
on the size of growing grains (see eq.~[\ref{eq:Ts}]),
it varies with time.
The temperature of the disk interior is much lower than 
that of the surface layer, as shown by Chiang \& Goldreich (1997).
The ratio of $T\sub{s}/T\sub{i}$ is about 4 in this case.
The temperature of the disk interior does not vary much with time
in spite of the substantial descent of the surface layer.
In other disk models examined below, we also found similar disk structure
to the standard disk.

\begin{figure}
\plotone{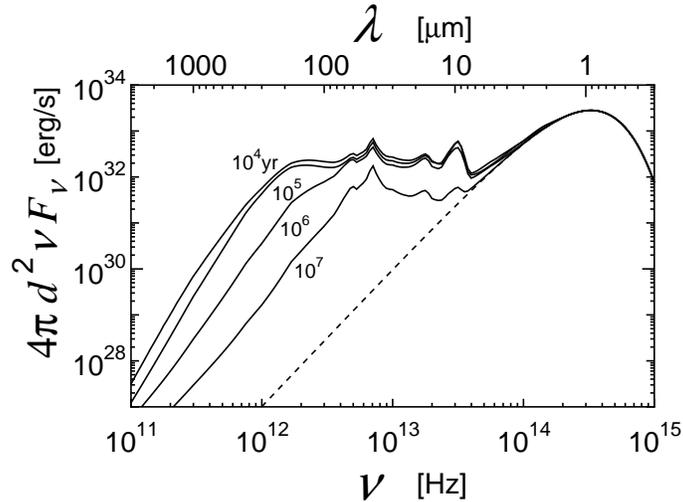}
\figcaption{Evolution of the SED of the standard disk.
The dashed line represents the black-body radiation from the central star.
Because of  grain growth and settling,
The energy fluxes from the disk decreases with time 
especially at long wavelengths of $\lambda \ge 100\mu$m.
\label{fig:10} }
\end{figure}

Using the obtained disk structure, we calculate the disk SED with 
Eqs.~(\ref{eq:Fs}) and (\ref{eq:Fi}).
We assume that the disk is face-on (the inclination angle $i$ is zero).
Figure~\ref{fig:10} shows evolution of the SEDs from the standard disk and 
the central 
star from $10^4$yr to $10^7$yr. The SED does not change much until $10^5$yr.
After that, the energy flux from the disk decreases  
especially at $\lambda \ge 100\mu$m.
The decreases in the disk optical thickness mainly reduces the flux
from the disk interior, which has a major contribution to the SED
at $\lambda \ge 100\mu$. The feature at 40$\mu$m, which comes from icy 
grains in the surface layer, does not decline much.
One the other hand, the silicate feature at 10$\mu$m depletes at $t=10^7$yr.
This is due to the ^^ ^^ inner hole" ($\lesssim 3AU$) of the disk optical 
thickness mentioned above.  
This result may explain the depletion of silicate feature 
in CoKu Tau/4 reported by Forrest et al.~(2004).
The disk life time estimated from the observed energy fluxes 
at millimeter/sub-millimeter wavelengths is about $10^{6-7}$yr 
(e.g., Wyatt at al. 2003). In our result on the standard disk, 
on the other hand,
the depletion time at the wavelengths is several$\times 10^{5}$yr.
We examine the possibilities to reconcile the
discrepancy below.

\begin{figure}
\epsscale{0.60}
\plotone{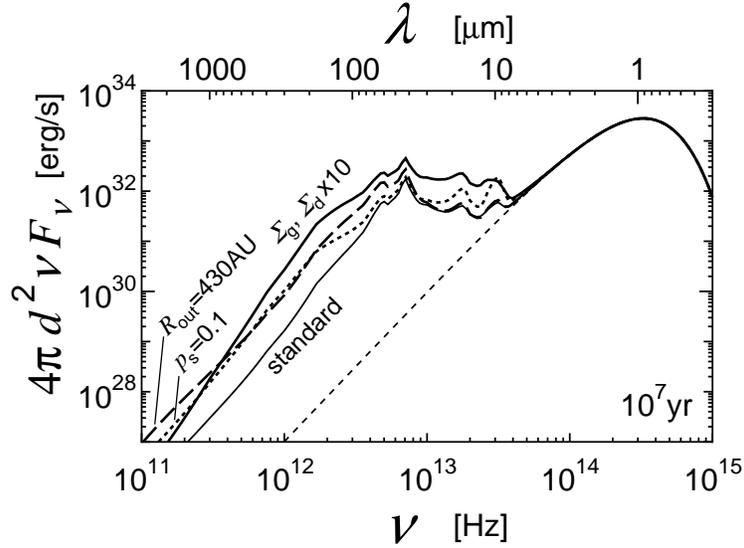}
\figcaption{Comparison in SEDs between various disks at $t=10^7$yr.
The solid line and the thick one correspond to the standard disk
and a disk ten times as massive as the standard one.
The dotted line and the long dashed line show the cases with a low sticking 
probability of $p\sub{s}=0.1$
and with a large disk outer radius of $r\sub{out}=430$AU, respectively.
\label{fig:11} }
\end{figure}

We performed three more similar simulations of grain growth and settling.
The first one is for a disk ten times as massive as the standard one,
the second one is for the standard disk but with the sticking probability 
$p\sub{s}=0.1$, and the third one is for a disk with the disk outer radius 
$R\sub{out}=430$AU. In Figure~\ref{fig:11}, we compare the SEDs in those cases 
with the standard disk at  $t=10^7$yr. In the case of the massive disk,
fluxes at sub-millimeter/millimeter wavelengths are larger than those 
in the standard disk. This is because of slow settling in the massive disk 
(see eq.~[\ref{eq:t_settle}]). The cases with a small sticking probability 
and with a large outer radius also result in larger fluxes at the long 
wavelengths. This is due to the slow grain growth at the outer disk radius
(see eq.~[\ref{eq:t_g}]).

\begin{figure}
\plotone{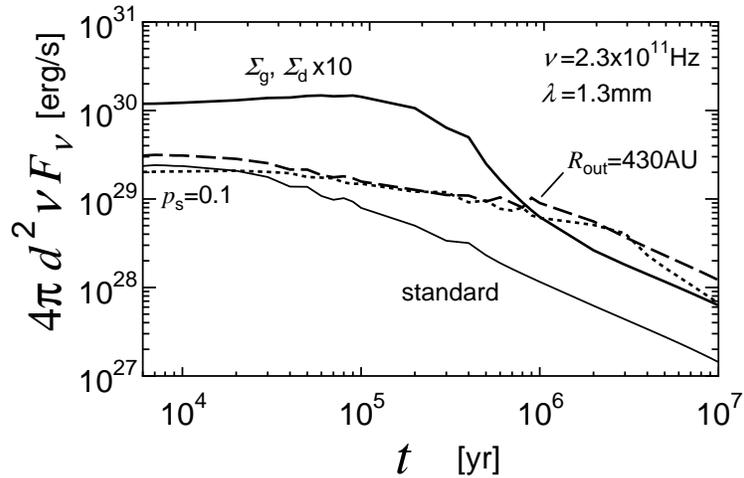}
\figcaption{Time-evolution of the energy fluxes at $\lambda=1$mm 
in the cases shown in Fig.~\ref{fig:11}.
The depletion time of the flux is longer
in the cases with a low sticking probability
and with a large outer disk radius than other cases.
\label{fig:12} }
\end{figure}

In Figure~\ref{fig:12}, the fluxes in these cases are displayed as a function of time.
The energy flux from the massive disk is larger than that from the standard
disk initially and starts to decrease at a few $10^5$yr similar to the 
standard disk.  In the cases with a small $p\sub{s}$
and with a large $R\sub{out}$, on the other hand, 
the flux starts to decrease more lately.
This is because of slow grain growth at the outer disk radii.
The depletion time in these cases is almost consistent with
that inferred from the SED observations.
Our results indicate that the decrease in the observed energy fluxes
at millimeter/sub-millimeter wavelengths can be explained by grain growth 
and settling without any disk depletion.
Hence the gaseous disks themselves can survive more than $10^7$yr.
In order to determine the disk life time, which is crucial for formation of 
gas giant planets, it is necessary to observe the decay of gas components.
\section{SUMMARY AND DISCUSSION}
\label{sec:summary}
We performed numerical simulations of dust growth and settling in 
passive disks and obtained the spatial and size distributions
of grains. As the first step of series of papers, we consider laminar disks
in the present paper.
Using the obtained grain distributions and the two-layer model, 
we also examined
evolution of disk structure and SEDs due to dust growth and settling.
Our results are summarized as follows:
\begin{enumerate}
\item
Evolution of grains in laminar disks is divided into the growth stage
and the subsequent settling stage. At the end of the growth stage, 
large grains settle to the mid-plane and form a dust layer, 
while small grains remain floating above the layer.
The size distribution, $N\sub{s}(a)$, of the floating grains is 
well-described by the power-law distribution analytically obtained
as $3/(2\pi p\sub{s} a^3)$ (eq.~[\ref{eq:size_dis}])
with the upper cut-off $a\sub{max}$ (eq.~[\ref{eq:a_max}]).
This size distribution is kept in the settling stage.
In the settling stage, relatively large floating grains settle earlier and
the upper cut-off radius of the size distribution decreases
as equation~(\ref{eq:a_max}).
The validity of equations~(\ref{eq:size_dis}) and (\ref{eq:a_max}) was 
confirmed with numerical simulations for wide parameter ranges.
The approximate solution of the spatial and size distribution,
$N(a,z)$, were obtained analytically, too (eq.~[\ref{eq:n-ns}]).
\item
The optical thickness of disks considerably decreases because of dust growth.
The optical thickness is governed by floating small grains
rather than large grains in the dust layer.
Rapid grain growth in the inner part of disks makes
the radial distribution of the disk optical thickness less steep
than that of the disk surface density $\varSigma$.
For disks with $\varSigma \propto R^{-3/2}$, the radial distribution
of the optical thickness is almost flat for all wavelengths
at $t \lesssim 10^6$yr.
The difference in the power-law index between the optical thickness
at $\lambda =1$mm and the surface density is generally about $1.5$.
This difference between the optical thickness and the surface density
should be taken into account
in analyses of observations with high-resolution, which will be
done by Spitzer Space Telescope or Atacama Large Millimeter Array (ALMA).
At $t \sim 10^7$yr, 
the optical thickness for infrared radiation is extremely small
in the inner region ($\lesssim $ 3AU).
This region may corresponds to inner holes observed by the Spitzer 
Space Telescope (Forrest at al. 2004; Meyer et al. 2004; Uchida et al. 2004).
\item
The surface layer directly irradiated from the central star descends 
because of grain settling and growth.
The descent of the surface layer is remarkable in an inner part of the disk
because grain growth due to Brownian motion depletes the micron-sized 
grains there.
The temperature of the disk interior, $T\sub{s}$, is much lower than
that of the surface layer, $T\sub{s}$, as shown by Chiang \& Goldreich (1997).
The ratio of $T\sub{s}/T\sub{i}$ is about 4 in our simulations.
\item
The energy flux from the disk decreases
due to grain growth and settling
especially at $\lambda \ge 100\mu$m.
In the standard disk with the sticking probability $p\sub{s}=1$
and the outer radius $R\sub{out}=$150AU, 
the depletion time of the energy fluxes at sub-millimeter/
millimeter wavelengths is obtained as several$\times 10^{5}$yr.
In the case with a small $p\sub{s}$ or with a larger $R\sub{out}$, 
the depletion time is longer. 
Our results indicate that
the decrease in the observed energy fluxes
at millimeter/sub-millimeter wavelengths
can be explained by grain growth and settling
without disk depletion.
The gaseous disks themselves can survive more than $10^7$yr.
The silicate feature at 10$\mu$m depletes on the time scale of $10^7$yr 
because of the formation of the extremely optical thin inner region.
\end{enumerate}

In the present paper, we consider only laminar disks.
Here, we comment on the effect of turbulence on dust growth and disk structure.
Vertical stirring by turbulence would prevent dust settling.
As shown by  Dullemond \& Dominik (2004b), vertical stirring by 
turbulence against dust settling is effective even in weakly turbulent 
disks with the parameter $\alpha =10^{-4}$ for the disk viscosity.
In turbulent disks, vertical stirring is expected to keep
the surface layer high compared with laminar disks.
On the other hand, turbulence also increases the relative velocity between 
grains. Since the collisional probability is proportional to the relative 
velocity, turbulence accelerates grain growth, too
(e.g., Weidenschilling 1984; Mizuno et al. 1988).
Acceleration of grain growth would result in a rapid decrease
in the opacity and the disk optical thickness.
This would make the surface layer lower.
It is not clear whether the turbulence makes the surface layer 
higher with stirring or lower with acceleration of grain growth.
Furthermore, in sufficiently strong turbulence,
collisional disruption due to high relative velocity creates
small grains and would increases the optical thickness.
To clarify which effect is dominant, it is necessary to perform
our simulations of grain growth and settling, by taking into account
above effects of turbulence.
In the next paper, we investigate cases of turbulent disks
and examine more realistic evolution of the optical thickness
and SEDs in protoplanetary disks.
%

We thank Y. Kitamura, M. Honda, and M. Momose for valuable comments.
We also thank the anonymous reviewer for many valuable comments and
suggestions. 
This work is supported by Grant-in-Aid of the Japanese Ministry of Education,
Science, and Culture (15740116 and 16340054)

\appendix
\section{BASIC EQUATIONS IN THE TWO-LAYER MODEL}
\label{sec:AppendixA}

Here we briefly explain the equations governing the structure 
of a passive disk in the two-layer model.
The optical thicknesses of the surface layer and the disk interior,
$\tau\sub{$\nu$,s}$, $\tau\sub{$\nu$,i}$ are given by
\begin{equation}
\tau\sub{$\nu$,s} 
= \int^\infty_{\tilde{z}\sub{s}}d\tilde{z} h_p \rho\sub{g} \kappa_\nu, \qquad
\tau\sub{$\nu$,i} 
= \int^{\tilde{z}\sub{s}}_{-\tilde{z}\sub{s}}d\tilde{z} h_p\rho\sub{g} \kappa_\nu,
\label{eq:tausi}
\end{equation}
respectively, where the height $\tilde{z}\sub{s}$ is the boundary between
the surface layer and the disk interior in the normalized vertical axis.
The opacity $\kappa_\nu$ is given by equation~(\ref{eq:kappa}) 
by the use of our numerical results. 
By the introduction of the normalized axis, the optical thicknesses of them
are independent of the disk temperature.
The Planck mean optical thicknesses of them are
\begin{equation}
\tau\sub{P,s}(T) = \frac{ \int^\infty_0 B_\nu(T)\tau\sub{$\nu$,s} d\nu}
                               {\int^\infty_0 B_\nu(T)d\nu},
\qquad
\tau\sub{P,i}(T) = \frac{ \int^\infty_0 B_\nu(T)\tau\sub{$\nu$,i} d\nu}
                               {\int^\infty_0 B_\nu(T)d\nu}.
\label{eq:taup}
\end{equation}
The boundary $\tilde{z}\sub{s}$ is indirectly determined by the equation
\begin{equation}
\tau\sub{P,s}(T_*) = \sin \theta,
\label{eq:zs}
\end{equation}
where $T_*$ is the effective temperature of the central star
and $\theta$ is the grazing angle (i.e. the angle between
the starlight and the disk surface).
By the use of $z\sub{s}=h_p\tilde{z}\sub{s}$, the grazing angle is given by
\begin{equation}
\theta = \arcsin \left(\frac{4R_*}{3\pi R}\right)
        +\arctan\left(\frac{z\sub{s}}{R}\frac{d\mbox{ln}z\sub{s}}{d\mbox{ln}R}\right)
        -\arctan\left(\frac{z\sub{s}}{R}\right),
\label{eq:angle}
\end{equation}
where $R_*$ is the radius of the central star.
When the disk optical thickness is small enough,
equation~(\ref{eq:zs}) does not have a positive solution of $\tilde{z}\sub{s}$.
This corresponds to the case with no disk interior.
In this case we set $\tilde{z}\sub{s}=0$ and, thus,
the optical thicknesses of the disk interior,
$\tau\sub{$\nu$,i}$ and $\tau\sub{P,i}(T)$, are zero. 

The grains in the surface layer have various temperatures
depending on their radii.
Instead of the various grain temperatures, 
we use the mean temperature of the surface layer $T\sub{s}$
defined by
\begin{equation}
T\sub{s}^4 = \frac{1}{4} 
\frac{\tau\sub{P,s}(T_*)}{\tau\sub{P,s}(T\sub{s})}
\left(\frac{R_*}{R}\right)^2 T_*^4.
\label{eq:Ts}
\end{equation}
The temperature of the disk interior $T\sub{i}$ is given by
\begin{equation}
T\sub{i}^4 = \frac{1}{2} \sin \theta \left(\frac{R_*}{R}\right)^2 T_*^4
\frac{1-\exp[-2\tau\sub{P,i}(T\sub{s})]}
     {1-\exp[-2\tau\sub{P,i}(T\sub{i})]}.
\label{eq:Ti}
\end{equation}
%
In equations~(\ref{eq:Ts}) and (\ref{eq:Ti}), 
we doubled the left-hand sides, 
assuming that the entire star is visible from the surface layer
(Dullemond et al. 2001).
Furthermore, 
we inserted the factor 2 into the exponential functions in
equation~(\ref{eq:Ti}) in order to 
take into account radiative transfer in the oblique direction in the disk 
interior. This is because, in the direction with the angle $\pi/3$ to the 
vertical axis, the optical thickness of the disk interior is double of that
in the vertical direction. 

Equations~(\ref{eq:zs})-(\ref{eq:Ti}) are coupled with one another.
We numerically solved these equations for a whole disk.
The derivative $dz\sub{s}/dR$ in equation~(\ref{eq:angle}) is evaluated
with the central difference using the both sides of grids
though the one-side difference is used at the innermost and the
outermost grids. 

By the use of the solutions to the above equations,
the emissions (or the energy fluxes) at a frequency $\nu$
from the surface layers and the disk interior,
$F\sub{$\nu$,s}$, $F\sub{$\nu$,i}$, are given by
\begin{equation}
F\sub{$\nu$,s} = \frac{1}{d^2}\int^{R_{out}}_{R_{in}}
B_\nu(T\sub{s}(R)) \tau\sub{$\nu$,s}
\left[1+\exp(-\tau\sub{$\nu$,i}/\cos i)\right]
2\pi R dR
\label{eq:Fs}
\end{equation}
and
\begin{equation}
F\sub{$\nu$,i} = \frac{\cos i}{d^2}\int^{R_{out}}_{R_{in}}
B_\nu(T\sub{i}(R)) 
\left[1-\exp(-\tau\sub{$\nu$,i}/\cos i)\right]
2\pi R dR,
\label{eq:Fi}
\end{equation}
respectively, where $d$ is the distance from the observer to the source
and $i$ is the inclination angle of the disk ($i=0$ means face-on).
In the above we assumed that $\tau\sub{$\nu$,s} \ll 1$.
The energy flux from the whole disk is given by their sum.

\newpage


\end{document}